\documentclass[aps,superscriptaddress,floatfix,11pt]{revtex4-1}

\usepackage[noend]{algpseudocode}
\usepackage[utf8]{inputenc}
\usepackage{algorithm}
\usepackage{graphicx}
\usepackage{hyperref}
\usepackage{amsmath}
\usepackage{amssymb}
\usepackage{fixmath}
\usepackage{xcolor}
\usepackage{bbold}
\usepackage{float}
\usepackage{amsthm}
\usepackage{soul,color} 

\newcommand{\PX}{\mathcal{P}}
\newcommand{\PME}{\mathcal{P_{ME}}}

\newcommand{\R}{\mathbb{R}}
\newcommand{\N}{\mathbb{N}}
\newcommand{\Rn}{{\R}^{N}}
\newcommand{\Rm}{{\R}^{M}}

\newcommand{\mymat}[1]{\mathbb{#1}}
\newcommand{\indvar}[1]{{#1}^{(i)}}
\newcommand{\depvar}[1]{{#1}^{(d)}}

\newcommand{\vol}[1]{|#1|}
\newcommand{\myvec}[1]{\mathbold{#1}}
\newcommand{\mean}[1]{\bar{#1}}
\newcommand{\myref}[1]{(\ref{#1})}
\newcommand{\Figure}[1]{\text{Figure}~\myref{#1}}
\newcommand{\Section}[1]{\text{Section}~\myref{#1}}
\newcommand{\Equation}[1]{\text{Equation}~\myref{#1}}
\newcommand{\Appendix}[1]{\text{Appendix}~\myref{#1}}

\newcommand{\Vcell}{\mathbb{V}}
\newcommand{\Pspace}{\mathbb{P}}
\newcommand{\Vcellvol}{\vol{\Vcell}}
\newcommand{\Vmean}{\mean{\Vcell}}

\newcommand{\vcell}{\myvec{v}}
\newcommand{\vcellscalar}{v}
\newcommand{\vmean}{\mean{\vcell}}
\newcommand{\vmeanscalar}{\mean{\vcellscalar}}

\newcommand{\rcell}{\myvec{r}}
\newcommand{\rjcell}{r_j}

\newcommand{\ajfwd}{a_j^+}
\newcommand{\ajbkwd}{a_j^-}

\newcommand{\rjfwd}{r_j^+}
\newcommand{\rjbkwd}{r_j^-}

\newcommand{\ucell}{\myvec{u}}
\newcommand{\ui}{u_i}
\newcommand{\uicell}{\ui}
\newcommand{\uimean}{\mean{\uicell}}

\newcommand{\uo}{u_o}
\newcommand{\uocell}{\uo}
\newcommand{\uomean}{\mean{\uo}}

\newcommand{\ug}{u_g}
\newcommand{\ugcell}{\ug}
\newcommand{\ugcellprime}{\ugcell^{\prime}}
\newcommand{\ugmean}{\mean{\ug}}

\newcommand{\zcell}{z}
\newcommand{\zcellprime}{\zcell^{\prime}}
\newcommand{\zmean}{\mean{\zcell}}

\newcommand{\lbv}{\myvec{lb}_\vcell}
\newcommand{\ubv}{\myvec{ub}_\vcell}
\newcommand{\lbr}{\myvec{lb}_\rcell}
\newcommand{\ubr}{\myvec{ub}_\rcell}
\newcommand{\lbu}{\myvec{lb}_\ucell}
\newcommand{\ubu}{\myvec{ub}_\ucell}
\newcommand{\lbz}{0}
\newcommand{\ubz}{ub_\zcell}

\newcommand{\ubg}{ub_g}

\newcommand{\ME}{ME}
\newcommand{\CBM}{CBM}
\newcommand{\FBA}{FBA}

\newcommand{\Ecoli}{E\ coli}

\newcommand{\cg}{c_g}
\newcommand{\ci}{c_i}
\newcommand{\sg}{s_g}

\newcommand{\betavec}{\myvec{\beta}}
\newcommand{\betaz}{\beta_z}
\newcommand{\betaug}{\beta_{\ug}}
\newcommand{\betaindvec}{\indvar{\betavec}}
\newcommand{\betadepvec}{\depvar{\betavec}}
\newcommand{\bprimevec}{\bvec^{\prime}}

\newcommand{\mmol}{mmol}
\newcommand{\mM}{mM}
\newcommand{\hour}{h}
\newcommand{\gCDW}{gCDW}
\newcommand{\flxunits}{\mmol \times \gCDW^{-1} \times \hour^{-1}}

\newcommand{\concunits}{\mM}
\newcommand{\Xunits}{\gCDW \times l^{-1}}
\newcommand{\Dunits}{\hour^{-1}}
\newcommand{\muunits}{\hour^{-1}}
\newcommand{\zunits}{\hour^{-1}}
\newcommand{\YXxunits}{\mmol \times \gCDW^{-1}}

\newcommand{\Kayser}{Kayser}
\newcommand{\Nanchen}{Nanchen}
\newcommand{\Folsom}{Folsom}

\newcommand{\identmat}{\mymat{1}}
\newcommand{\Sij}{S_{ij}}
\newcommand{\Gmat}{\mymat{G}}
\newcommand{\Smat}{\mymat{S}}

\newcommand{\vscalar}{v}
\newcommand{\vvec}{\myvec{\vscalar}}
\newcommand{\vdepvec}{\depvar{\vvec}}

\newcommand{\vindvec}{\indvar{\vvec}}

\newcommand{\cvec}{\myvec{c}}
\newcommand{\bvec}{\myvec{b}}
\newcommand{\lbvec}{\myvec{lb}}

\newcommand{\ubvec}{\myvec{ub}}

\newcommand{\Pvbb}[2]{\PX(v~|~#1, #2)}
\newcommand{\Pexact}{\PX_{\psi}}
\newcommand{\Papprox}{\PX_{\phi}}

\newcommand{\avec}{\myvec{a}}
\newcommand{\aindvec}{\indvar{\avec}}

\newcommand{\adepvec}{\depvar{\avec}}

\newcommand{\dscalar}{d}
\newcommand{\dvec}{\myvec{\dscalar}}
\newcommand{\dindvec}{\indvar{\myvec{\dscalar}}}
\newcommand{\ddepvec}{\depvar{\myvec{\dscalar}}}
\newcommand{\Dmat}{\mymat{D}}
\newcommand{\Dindmat}{\indvar{\Dmat}}

\newcommand{\Ddepmat}{\depvar{\Dmat}}

\newcommand{\vmeanindvec}{\indvar{\vmean}}

\newcommand{\vmeandepvec}{\depvar{\vmean}}

\newcommand{\Sigmamat}{\mymat{\Sigma}}
\newcommand{\Sigmaindmat}{\indvar{\Sigmamat}}

\newcommand{\Sigmadepmat}{\depvar{\Sigmamat}}
\newcommand{\ascalar}{a}

\newcommand{\notimplies}{\;\not\nobreak\!\!\!\!\implies}
\newcommand{\eg}{\emph{e.g.}}
\newcommand{\ie}{\emph{i.e.}}
\newcommand{\etc}{\emph{etc.}}

\newcommand{\cgDX}{\cg D / X}

\newcommand{\zeq}{\zmean \approx D}
\newcommand{\ugineq}{\ugmean \lesssim \cgDX}
\newcommand{\ugeq}{\ugmean \approx \cgDX}

\newcommand{\betavz}{\beta_{z}^{v}}
\newcommand{\betavug}{\beta_{\ug}^{v}}

\begin{document}

\title{%
	Inferring metabolic fluxes in nutrient-limited continuous cultures: A Maximum Entropy Approach with minimum information
}

\author{Jose A. Pereiro-Morejón}
\affiliation{%
	Group of Complex Systems and Statistical Physics. Physics Faculty, University of Havana, CP 10400. La Habana, Cuba and \\%
	Biology Faculty, University of Havana, CP 10400. La Habana, Cuba
}
\author{Jorge Fern\'andez-de-Cossio-D\'{\i}az}
\affiliation{%
	Laboratory of Physics of the Ecole Normale Supérieure, CNRS UMR 8023 \& PSL Research, Paris, France
}
\author{Roberto Mulet}
\email{mulet@fisica.uh.cu}
\affiliation{%
	Group of Complex Systems and Statistical Physics.
	Department of Theoretical Physics, Physics Faculty, University of Havana, Cuba
}

\begin{abstract}
	We propose a new scheme to infer the metabolic fluxes of cell cultures in a chemostat.
	Our approach is based on the Maximum Entropy Principle and exploits the understanding of the chemostat dynamics and its connection with the actual metabolism of cells.
	We show that, in continuous cultures with limiting nutrients, the inference can be done with {\it limited information about the culture}: the dilution rate of the chemostat, the concentration in the feed media of the limiting nutrient and the cell concentration at steady state.
	Also, we remark that our technique provides information, not only about the mean values of the fluxes in the culture, but also its heterogeneity.
	We first present these results studying a computational model of a chemostat.
	Having control of this model we can test precisely the quality of the inference, and also unveil the mechanisms behind the success of our approach.
	Then, we apply our method to E. coli experimental data from the literature and show that it outperforms alternative formulations that rest on a Flux Balance Analysis framework.
\end{abstract}

\maketitle

\section{Introduction}
\label{sec:Intro}
	 
The study of cellular metabolism is a research field with a direct impact on the biotechnological industry.
Indeed, cell culture-derived products are a major part of a multi-billion market~\cite{weng2020reduction}.
These products are obtained by exploiting the capabilities of cellular metabolism to produce molecules with a wide range of chemical complexity.
Cells are cultivated in three common modes: batch, fed-batch and continuous~\cite{xu2017bioreactor}.
In batch, cultivation starts with a medium rich in nutrients that is consumed by the cells, often until starvation.
Similarly, fed-batch cultures start with a nutrient pool, which is resupplied in discrete time intervals, maintaining the cells alive for longer periods of time.
On the other hand, in continuous mode, fresh medium constantly replaces culture fluid at a given rate~\cite{ozturkEngineeringChallengesHigh1996}.

The chemostat is a prototypical continuous cultivation device developed in the 50s~\cite{monodGrowthBacterialCultures1949, novickDescriptionChemostat1950}.
Chemostats are often operated at constant volume and in steady state, which is reached when macroscopic variables of the culture stay constant in time (basically cell and extracellular metabolite concentrations).
It is also common to specify which medium component is limiting cell growth.
Although the advantages of continuous cell culture have been widely discussed in the literature~\cite{wernerSafetyEconomicAspects1992, griffithsAnimalCellCulture1992, kadouriMythsMessagesConcerning1997, wernerLetterEditor1998, croughanFutureIndustrialBioprocessing2015}, the use of these techniques over batch or fed-batch is hampered by the complexity of continuous systems, \ie~culture heterogeneity, hysteresis, multi-stability or sharp transitions between metabolic states~\cite{mulukutlaMultiplicitySteadyStates2015, europaMultipleSteadyStates2000, cAnalysisCHOCells2001a, hayterGlucoseLimitedChemostat1992, gambhirAnalysisCellularMetabolism2003, follstadMetabolicFluxAnalysis1999, fernandez-de-cossio-diazCharacterizingSteadyStates2017}.
This complexity negatively impacts the yield of bio-processes.
In particular, culture heterogeneity is estimated to generate losses of more than 30\% in industrial-scale fermentation~\cite{fernandesExperimentalMethodsModeling2011}.

Culture performance is an emergent property	derived from the individual metabolic state of each cell~\cite{gonzalez-cabaleiroHeterogeneityPureMicrobial2017}, but also the result of interactions between cells.
It is fundamental to connect metabolic states at the individual cell level, to macroscopic properties at the culture level.
This connection can guide efforts to understand cellular metabolism in a continuous regime and suggest strategies to improve production efficiency~\cite{perez-fernandezInsilicoMediaOptimization}.

In this task, the community has been assisted by an increasing number of accurate experimental techniques that generate large amounts of data.
In particular, information about cellular metabolism, at the level of individual reactions, has led to the development of genome-scale metabolic networks ($GEM$s)~\cite{kanehisaDataInformationKnowledge2014,caspiMetaCycDatabaseMetabolic2016, palssonSystemBiologyConstraintbased2015}.
Although at present, a full characterization of cellular metabolism is not feasible, a Constraint-Based Modeling (CBM) approach helps to integrate a variety of data types ($\eg$ stoichiometric, thermodynamic, dynamic, genetic, \etc) that restrict as much as possible the space of feasible phenotypes that the metabolic network can display.
	
Constraint-based methods such as Flux Balance Analysis ($\FBA$)	have been extensively used to predict a wide range of metabolic observables ($\eg$ culture growth rate, $ATP$ production, \etc), especially for bacterial batch cultures in the exponential growth phase~\cite{ibarraEscherichiaColiK122002, palssonSystemBiologyPropereties2006, schuetzSystematicEvaluationObjective2007, zengModellingOverflowMetabolism2019}.
$\FBA$ can also be exploited in combination with experimental data, if the latter provides only a partial knowledge about macroscopic properties of the culture.
For example, if the growth rate or metabolic concentrations are known from experimental data, this information can be introduced in the $\FBA$ framework to refine predictions about other fluxes in the network~\cite{robinsonAtlasHumanMetabolism2020, rivas-astrozaMetabolicFluxConfiguration2020}.
However, as we will discuss in more detail below, typical $\FBA$ formulations can hardly provide any insights about important culture properties such as cellular heterogeneity.

A more general methodology that exploits the Maximum Entropy Principle $(\ME)$~\cite{jaynesInformationTheoryStatistical1957} combined with a constraint-based model, in order to formulate a probabilistic description of the metabolic state of cells in the culture, has been used recently~\cite{muntoniRelationshipFitnessHeterogeneity2021, demartinoStatisticalMechanicsMetabolic2018,demartinoIntroductionMaximumEntropy2018,demartinoGrowthEntropyBacterial2016}.
In particular, it has been shown that $\ME$ distributions provide a better fit to measured flux observables than plain $\FBA$ models \cite{demartinoStatisticalMechanicsMetabolic2018}.
Also, $\ME$-derived growth rate distributions have been compared to experiments with good results, using single-cell data at different sub-inhibitory antibiotic concentrations~\cite{demartinoStatisticalMechanicsMetabolic2018}.
A more recent formulation exploits the availability of experimental data from the culture, to further limit the available solution space for the metabolism \cite{muntoniRelationshipFitnessHeterogeneity2021}.
In practice, the authors of \cite{muntoniRelationshipFitnessHeterogeneity2021} not only fix the growth rate, but all the experimental fluxes.
This approach however, leaves open relevant questions: i) What is the minimum number of fluxes that we need to fix to have a proper description of the system? ii) Is it preferable to fix some particular fluxes instead of others? If we fix too many fluxes, we leave too few of them to test the predictions derived from the inference.
Moreover, we also increase the risk of introducing in the inference process unnecessary biases that may come from uncontrolled experimental errors made in the measurement process.

The main goal of this work is to provide a novel $\ME$ scheme to infer the flux space of Genome Scale Metabolic networks grown in nutrient-limited continuous cell cultures, {\it using a minimum set of experimental data}.
To reach this goal we exploit a constraint-based model for a continuous culture introduced in reference~\cite{fernandez-de-cossio-diazCharacterizingSteadyStates2017}.
It provides a detailed characterization of the steady state of the chemostat, coupling cell metabolism with the dynamics of extracellular observables.
Using the Maximum Entropy Principle in reference~\cite{fernandez-de-cossio-diazMaximumEntropyPopulation2019} we already showed that one can observe non-trivial distributions for metabolic fluxes, supporting the heterogeneity of the culture in the chemostat.
Here we go one step further, closing the gap between simulated models and real experimental data.
We support the idea that with the knowledge of only external parameters: the chemostat dilution rate, cellular concentration at steady state, and the concentration of the limiting metabolite in the feed medium; we can obtain a description of the metabolism in a continuous culture.

The rest of the work is organized as follows.
In the next section we introduce the main concepts of constraint modeling techniques for the metabolism and how they can be applied to continuous cultures with limiting nutrients.
Then, we introduce Flux Balance Analysis and the Maximum Entropy Principle in section \ref{sec:Infer}.
There, we explain how they can be used to infer the metabolic fluxes using experimental data from this kind of cultures.
Later, in section \ref{sec:Results}, we present the results of our work.
We first make an analysis of the consequences of imposing different constraints in $\FBA$ and $\ME$.
Then, we exploit the $\ME$ method to infer the metabolic state of a simulated chemostat culture using a simple model of the cellular metabolism.
Finally, we show the application of our $\ME$ formulation on a genome-scale network inferring a set of literature-available experimental flux observables from glucose-limited $\Ecoli$ chemostat cultures.
For completeness, we compare and discuss the results obtained with our methodology ($\ME$) with the solutions obtained through different $\FBA$ approximations.

\section{Constraint-Based Metabolic Models}
\label{sec:CBM}

The formulation of models able to describe, from first principles, the evolution and properties of biological networks (such as $GEM$s) is in general an open problem.
Among the limitations there are the complexity of interactions between its many components, the large number of parameters (usually prohibitively large) required to formulate a complete description of the system, and the fact that they are subject to evolution.
The latter means that the models need to be continuously updated~\cite{palssonSystemBiologyConstraintbased2015}.
Therefore, it is common to study the metabolism considering only the known effect of constraints over the possible physiological states of the system.
These constraints can be physicochemical, spatial, topological, environmental, or regulatory in nature.
This approach, called Constraint-Based Modeling ($CBM$), leads to the formulation of solution spaces rather than the computation of a single solution~\cite{palssonSystemBiologyConstraintbased2015}.

\subsection{Metabolic Networks and Constraint Based Modeling (CBM)}
\label{sec:GEM-CBM}

A metabolic network is built connecting metabolites as described by the stoichiometry of the reactions in the cell.
If the network includes a significant portion of the known chemical reactions comprehended in the organism genome, it is called a Genome-Scale Metabolic Network.
It constitutes the basis to formulate a constraint-based model of cellular metabolism~\cite{guCurrentStatusApplications2019} where the rate of change of the concentration of any metabolite depends on the combined effect of all reactions that involve it.

For a network with $N$ reactions and $M$ metabolites, a balance equation can be written as:
\begin{equation}
	\frac{dm_i}{dt} = \sum_j^N \Sij ~ v_j
	\label{eq:dm_dt}
\end{equation}
where $1\le i\le M$, $1\le j\le N$, $m_i$ is the intracellular concentration of metabolite $i$, $v_j$ is the flux value assigned to reaction $j$, and $S \in \R^{M \times N}$ is the stoichiometric matrix where $\Sij$ is the stoichiometric coefficient of metabolite $i$ in reaction $j$.
The common convention is that $\Sij=0$ means that the metabolite does not participate in the reaction, $\Sij<0$ that the metabolite participates as a reactant, and $\Sij>0$ that it participates as a product.
The information required to model the time dependency of $v_j$	is not commonly available.
Therefore, it is usual to introduce a quasi-steady state assumption for intracellular metabolites~\cite{palssonSystemBiologyConstraintbased2015}, that separates the time scales in the system and allows writing equation~\myref{eq:dm_dt} as:
\begin{equation}
	0 = \sum_j^N \Sij v_j
	\label{eq:m_i_balance}
\end{equation}
A particular flux configuration is specified by the vector $\vcell\in\Rn$ of all flux values $v_j$ included in the network.
In practice, besides the biochemical reactions ($\eg$ catalyzed by enzymes), this vector may contain additional reactions (often artificial) that are included according to a variety of modeling reasons.
An example are exchange reactions ($\ucell \in \Rm$), which model the transport of metabolites between the cell and its environment.
Another important component of $\vcell$ is the biomass reaction ($z \in \R$), which represents the synthesis of new biomass (and secondary products) from a set of precursors.
The exchanges and the biomass reaction, represent the boundary of the system.
The rest are considered to be internal reactions ($\rcell \in \R^{N-M-1}$), in short $\vcell \equiv \{\ucell,~\rcell,~\zcell\}$.

\Equation{eq:m_i_balance} constitute the first set of constraints that restricts the flux configurations of the network.
They form a linear system of equations.
Any solution is a vector $\vcell$ that satisfies the balance of mass for each metabolite.
However, a typical network has more reactions than metabolites $(M < N)$, which leads to fewer constraints than variables (fluxes)~\cite{palssonSystemBiologyConstraintbased2015}.
The system is then under-determined.
An infinite set of vectors $\vcell$ satisfies the system of equations.

These constraints lead to an unbounded solution space.
Therefore it is common to add a set of inequalities to impose bounds on $\vcell$, such as:
\begin{gather}
   \lbr \le \rcell \le \ubr \label{eq:v_j_bounds} \\
   \lbu \le \ucell \le \ubu \nonumber \\
   \lbz \le \zcell \le \ubz \nonumber
\end{gather}
where $\lbr$ and $\ubr$ are the lower and upper bounds of the internal reactions, which typically contain information about thermodynamic irreversibility and catalytic capacity.
On the other hand, $\lbu$ and $\ubu$ are the bounds of the exchange reactions controlling the metabolites that the network can consume or produce, and are linked to properties of the cell membrane ($\eg$~the presence transporters, ion channels, \etc).
Finally, the biomass reaction can be upper bounded by $\ubz$, if necessary.

In general, any new information about the culture is integrated by adding new balance-like equations (\eg~\Equation{eq:m_i_balance}) or changing the bounds of the reaction fluxes (\eg~\Equation{eq:v_j_bounds})~\cite{orthWhatFluxBalance2010}.
For example, we can define a new constraint that accounts for physical and spatial restrictions resulting from the limited resources accessible to the cell ($\eg$ cell volume, membrane area, enzyme solubility, proteome, \etc)~\cite{begIntracellularCrowdingDefines2007,fernandez2017limits,fernandez2018physical,scott2010interdependence,basan2015overflow}.
It can be formulated as:
\begin{equation}
	\sum_j^N (\ajfwd \rjfwd + \ajbkwd \rjbkwd) \le 1
	\label{eq:cost}
\end{equation}
where each internal reaction $\rjcell$ in the network is split into its forward and backward component such that $\rjcell = \rjfwd - \rjbkwd$ and $\rjfwd, \rjbkwd \ge 0$ where $\ajfwd$ and $\ajbkwd$ are normalized cost coefficients associated with each component of the reaction~$j$ respectively.
From the mathematical point of view, it is important to note that these constraints define a convex space of feasible flux configurations~\cite{boydConvexOptimization2004}.

\subsection{Constraint-Based Modeling of the chemostat}
\label{sec:ChCBM}
	
In references~\cite{fernandez-de-cossio-diazCharacterizingSteadyStates2017, fernandez-de-cossio-diazMaximumEntropyPopulation2019} we developed a constraint-based model of genome-scale metabolic networks coupled to the dynamical equations governing a chemostat.
Here we go a step further, linking culture observables with the constraints that affect the metabolic spaces at steady state.

\begin{figure}
	\centering
	\includegraphics[scale = 0.20]{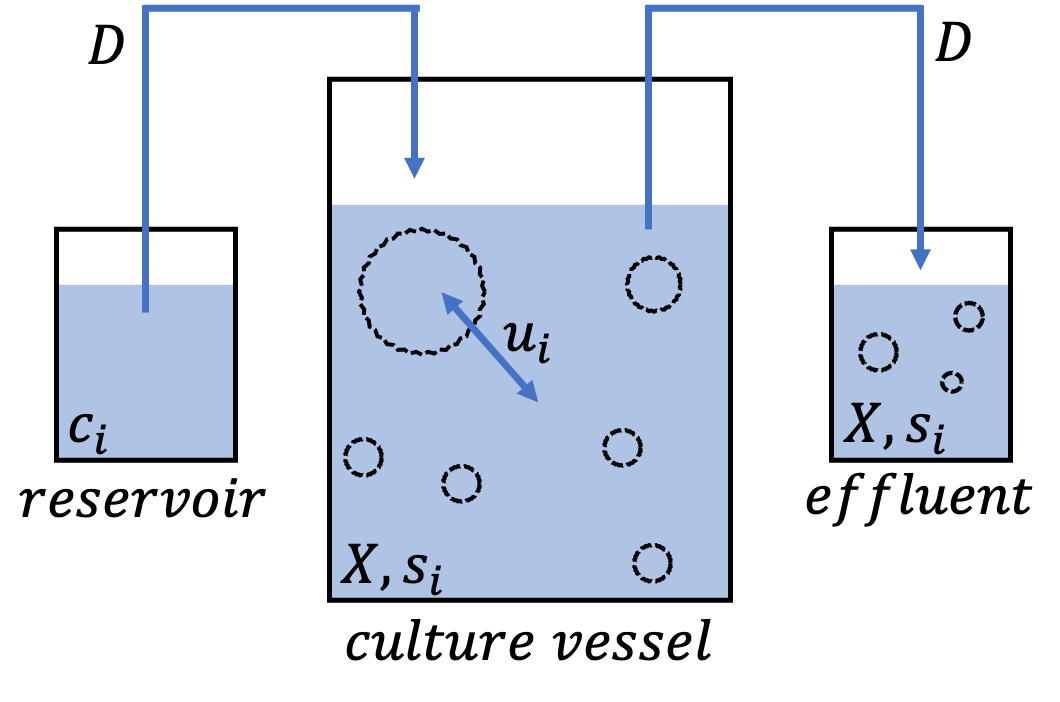}
	\caption{
		Schema of a chemostat. Fluxes of matter are indicated by an arrow.
		The most important chemostat parameters are listed:
		dilution rate $D$, cell concentration $X$,
		exchange of a metabolite between the cells and the medium $\uicell$, and
		the concentration of a metabolite in the feed medium $c_i$ and in the culture vessel $s_i$.
		Adapted from~\cite{fernandez-de-cossio-diazCharacterizingSteadyStates2017}.
	}
	\label{fig:ch}
\end{figure}
	 
In a chemostat, a cell culture is maintained in a continuous regime where fresh medium is pumped into the culture vessel at the same rate that it is extracted, such that the working volume remains constant~\cite{benyahiaMacroscopicModelingMammalian2015}.
In \Figure{fig:ch} we present a schematic picture of the chemostat.
The dynamics of cell and metabolite concentrations in the vessel, $X$ ($\Xunits$) and $s_i$ ($\concunits$), for a classic well-mixed chemostat with a single species, can be expressed as~\cite{fernandez-de-cossio-diazCharacterizingSteadyStates2017}:
\begin{gather}
	\frac{dX}{dt} = (\mu - D)X 
	\label{eq:CH-dX_dt} \\
	\frac{ds_i}{dt} = - \uimean X + (c_i - s_i)D
	\label{eq:CH-dsi_dt}
\end{gather}
where $D$ ($\Dunits$) is the dilution rate, $\mu$ ($\muunits$) is
the observable culture growth rate, $c_i$ ($\concunits$) and $\uimean$~($\flxunits$)
are the concentration in the fresh medium and the observable exchange rate
of metabolite $i$, respectively.
Here and in what follows we will use an overbar (as in $\uimean$) to distinguish the average value of a flux across all the cells in the culture, from its value in single cells ($u_i$).
\Equation{eq:CH-dX_dt} says that the rate of change of $X$ is determined by the culture growth rate and its elimination due to medium exchange.
Similarly, \Equation{eq:CH-dsi_dt} reflects that the rate of change of any metabolite concentration in the vessel is a balance between its average exchange with the cells
(a positive $\uicell$ means uptake) and how much of it is being pumped in and out of the vessel.
	
The culture growth rate (usually the relevant observable) $\mu$, can be modeled to include any	metabolic process that impacts the average growth rate of the culture ($\eg$ toxicity, cellular death rate, \etc).
In this work we only consider the biomass production rate $\zcell$ ($\zunits$), so:
\begin{equation}
	\mu = \zmean \nonumber
\end{equation}
where, as mentioned before, $\zcell$ is just a component of the flux vector $\vcell$ and $\zmean$ is its average value on the cell population.
It models the flux requirement for cellular division.

The contribution of the negative terms in equations~\myref{eq:CH-dX_dt} and~\myref{eq:CH-dsi_dt} (right-hand side) enables the possibility of a steady state regime.
One of the main applications of the chemostat is that cultivation can be sustained for a long time in a constant environment.
We exploit this fact to decouple the cellular physiology from extra-cellular dynamical processes.
This is a major difference with batch cultures, where cells are in a constantly changing environment~\cite{smithTheoryChemostatDynamics1995}.
Therefore, for a chemostat in steady state, two new constraints can be derived from equations~\myref{eq:CH-dX_dt} and~\myref{eq:CH-dsi_dt}:
\begin{gather}
	\uimean \le c_i D / X
	\label{eq:Ch_ui_stst_ineq_constraints} \\
	\zmean = \mu = D
	\label{eq:Ch_mu_stst_constraints}
\end{gather}
The first equation \eqref{eq:Ch_ui_stst_ineq_constraints} simply states that $s_i\ge0$ in steady state \cite{fernandez-de-cossio-diazCharacterizingSteadyStates2017}.

With this, the full stack of equations needed to define the constraint-based model for a chemostat, in steady state, is complete.
The set of equations~\myref{eq:m_i_balance})-(\ref{eq:cost} reflect the metabolic constraints within each cell, and equations~\myref{eq:Ch_ui_stst_ineq_constraints} and~\myref{eq:Ch_mu_stst_constraints} constraint the average values of the fluxes in the culture.

\subsection{Metabolic Flux Spaces}
\label{sec:MetSpaces}
	
In the previous sub-sections, we presented two types of constraints.
Depending on the source of the information encoded, they can restrict the flux configurations at a single cell level or at a culture level.
For instance, if a metabolic reaction is considered to be thermodynamically irreversible, a constraint enforcing such behavior must be applied to all cells in the culture.
On the other hand, data derived from experimental measurements made at a population level ($\eg$ culture growth rate), are interpreted differently: as the average over the configurations of all cells in the culture.
But these latter constrains do not necessarily imply that a particular restriction must be fulfilled at a single cell level.
For example, if a culture is considered to be growing at a given rate, this does not imply that all cells are growing at such speed.
Some cells might be growing faster, and some slower: it is the population average what defines the measured value.

Therefore, it is convenient to introduce two separated spaces: $\Vcell$ as the space of all feasible flux configuration $\vcell$ that a particular cell metabolism can display, and $\Vmean$ as the space of all feasible average flux configurations $\vmean$ compatible with the measured observables of the culture.
If the nature of all constraints acting over the system is linear, as it is in our case, both $\Vcell$ and $\Vmean$ are high-dimensional convex polytopes~\cite{boydConvexOptimization2004}.
The convexity of $\Vcell$ implies that $\Vmean \subseteq \Vcell$.
That is, any observable feasible flux configuration $\vmean$ at the population level, is a feasible flux configuration for single cells $\vcell$, and no unfeasible single-cell flux configuration can be observed at the population level.

The clear distinction between both spaces, $\Vcell$ and $\Vmean$, plays a key role in the definitions of the inference models in the next sections.
In our context, the final formulation for these spaces can be written as:
\begin{gather}
	\Vcell = \{
		\vcell ~|~
		\Smat \vcell = \myvec{0};~
		\vcell \in [\lbv, \ubv] ;~
		\sum_j^N (\ajfwd\rjfwd + \ajbkwd\rjbkwd) \le 1;~
		\rjfwd, \rjbkwd \ge 0 ;~
		\rjfwd - \rjbkwd = \rjcell \in \vcell 
	\} \nonumber \\
	\Vmean = \{
		\vmean \in \Vcell ~|~
		\uimean \le c_i D / X ;~ 
		\zmean = D ;~
		\uimean, \zmean \in \vmean
	\} \nonumber
\end{gather}
%

An important remark about the definition of $\Vcell$ is that it is independent of the chemostat dynamics, it is a property of the cells.
The environmental conditions are taken into account only in the definition of $\Vmean$ through equations~\myref{eq:Ch_ui_stst_ineq_constraints} and~\myref{eq:Ch_mu_stst_constraints}.
Moreover, since usually in a chemostat, $D$ and $\cvec$ (the feed medium composition) are controlled by the researcher, $X$ alone encodes all the information dependent on the chemostat dynamics.

\subsection{Nutrient-limited cultures}
\label{sec:nutrient_limited_cultures}
	
As we already mentioned we will focus our attention on chemostat cultures with a known limiting nutrient~\cite{smithTheoryChemostatDynamics1995}.
In practice, this means that from all the exchange constraints defined in equation~\myref{eq:Ch_ui_stst_ineq_constraints}, only one, associated with this limiting nutrient, is constraining the network.
In all the experiments presented here, glucose is the limiting nutrient.
That is, the only constraints affecting $\Vmean$ are $\zmean = D$ and $\ugmean \le \cgDX$, where $\ugmean$ is the observable uptake rate of the limiting nutrient (glucose) and $\cg$ is the concentration of this nutrient in the feed medium.
	
The rest of the constraints over $\Vmean$ are considered to be non-restrictive and therefore do not influence the culture.
In this context, we can build a simpler definition of~$\Vmean$:
\begin{gather}
	\Vmean = \{
			\vmean \in \Vcell ~|~ 
			\ugmean \le \cgDX ;~ \zmean = D ;~
			\ugmean, \zmean \in \vmean
		\} \nonumber
\end{gather}
%

\begin{figure}
	  \centering
	  \includegraphics[scale = 0.25]{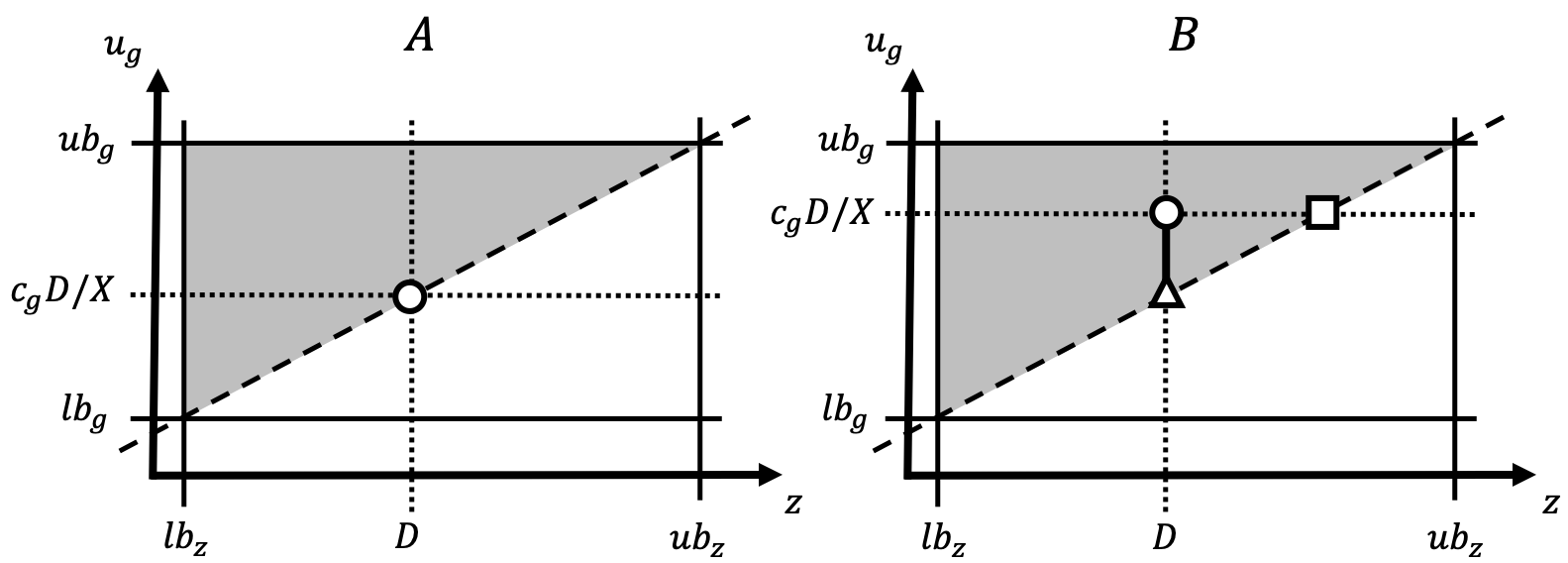}
	  \caption{Projections of $\Vcell$ and $\Vmean$ on the 2D plane $(\zcell,\ug)$ for a chemostat in steady state. The solid and dashed lines represent the constraints defining $\Vcell$, which is shown in gray. The dotted lines indicate the constraints defining $\Vmean$.
	  The left panel (A) shows a situation where the exchange bound is so tight that the projection of $\Vmean$ on the $(\zcell,\ug)$ plane is reduced to a single point (white circle).
      Meanwhile, in the right panel (B), the projection of $\Vmean$ on the $(\zcell,\ug)$ is a line (solid vertical black segment).
	  In both cases, the circle marks a point $\vmean$ where the culture is glucose limited.
	  See the main text for the meaning of the square and triangle markers in (B).}
	  \label{fig:pol_stst_chemostat}
\end{figure}

Now, the chemostat constraints are affecting $\Vmean$ only in the $(\zmean,\ugmean)$ subspace.
\Figure{fig:pol_stst_chemostat} shows a schematic representation of this subspace and the two typical scenarios that can occur in a nutrient-limited culture.
In the horizontal axis we plot the growth rate of the cell, and in the vertical axis the glucose consumption rate.
The shadowed area represents the projection of $\Vcell$, the space of feasible flux configurations, on the plane: no cell can display a $(z,\ug)$ pair outside of this area.
Moreover, since in a chemostat at steady state the average growth rate of the culture is set by the dilution rate $D$, all the possible solutions should be consistent with distributions where $\zmean=D$ (see vertical dotted lines on both panels of the figure).
This reduces the possible degeneracy of $\Vmean$ in this subspace only to the $\ugcell$ dimension, where the average consumption rate should be lower than $\cgDX$ (\ie~the system is restricted to those distributions where $\ugmean$ rests below the horizontal dotted line in the panels).

Such a combination of constraints leads to two typical scenarios.
One is represented in Panel A, where the size of $\Vmean$ is reduced to the minimum volume allowed by the environmental constraints (equations~\myref{eq:Ch_ui_stst_ineq_constraints} and~\myref{eq:Ch_mu_stst_constraints}).
In the $(z,\ug)$ plane, $\Vmean$ is reduced to a single point (white circle in the figure).
In these conditions, the culture is growing with the maximum possible $\zmean$/$\ugmean$ yield and larger values of $X$ are not feasible given the nutrient feed rate ($\cg D$) and the definition of $\Vcell$.
We stress that, although $\Vmean$ is determined in the ($z$,~$\ug$) plane in this example, that does not imply that $\Vmean$ is not degenerated in other dimensions.
The other scenario is represented in Panel B of \Figure{fig:pol_stst_chemostat}.
In this case $X$ is not optimal, and we have a degenerated $\Vmean$ even in the ($z$,~$\ug$) subspace (continuous vertical line).
The major difference between both scenarios is that in Panel A, the culture is using the full carrying capacity of the medium, and in Panel B it is not.

\section{Inference of the Metabolic State}
\label{sec:Infer}

Most constraint-based frameworks consist of two stages: I)~the specification of the constraints and the definition of the feasible spaces, and II)~the methods to formulate a description of the metabolism from them~\cite{bordbarConstraintbasedModelsPredict2014}.
In the previous section we already discussed point I, here we focus the attention on the second step presenting two standard approaches.

\subsection{Flux Balance Analysis}
\label{sec:FBA}

Flux Balance Analysis ($\FBA$) is a widely used methodology that addresses the typical degeneration of the metabolic solution space by choosing an objective function ($f$) (or a stack of them) that the cell metabolism ``optimizes''.
This assumption is not necessarily based on experimental data, but it is an educated guess about the evolutionary pressures to which the biological system is exposed~\cite{orthWhatFluxBalance2010}.

$\FBA$ has been applied, for several decades now, to model cell cultures at optimal growth conditions with remarkable results~\cite{varmaStoichiometricFluxBalance1994, garciasanchezComparisonAnalysisObjective2014a, lewisOmicDataEvolved2010, moralesValidationFBAModel2014}.
A very popular formulation for bacterial cultures is to set the objective function equal to the biomass production rate, $z$.
This is particularly justified in rich medium batch cultures, during the exponential growth phase, where the fastest growing cells dominate the culture population.
In these circumstances, $\FBA$ has proven to predict the growth rate of $\Ecoli$ cultures~\cite{ibarraEscherichiaColiK122002} for single carbon source conditions, and even the priority of nutrient uptakes for more complex mediums~\cite{begIntracellularCrowdingDefines2007}.

From a computational point of view, $\FBA$ has the advantage that, if the proposed objective function is formulated as a linear function over a convex space, the optimum flux configurations can be found efficiently using Linear Programming~\cite{lloydCOBRAmeComputationalFramework2018}.
Typically, $\FBA$ formulations~\cite{ibarraEscherichiaColiK122002, bordbarConstraintbasedModelsPredict2014, schuetzSystematicEvaluationObjective2007, herrmannFluxSamplingPowerful2019} do not make the explicit distinction between population and single-cell level metabolic spaces.
All constraints are applied over a unique space.
In the case of our two spaces formalism this is equivalent to making $\Vcell \equiv \Vmean$, which hides an implicit culture homogeneity assumption.
This can be justified because generally, the goal is just to infer an observable flux configuration $\vmean$ which optimizes the objective function and a convex $\Vmean$ is not affected by such an assumption.
The problem to solve can then be stated as:
\begin{gather}
	\text{optimize arg}_{\vmean} ~ f(\vmean)
	\label{eq:FBA_formulation} \\
	\text{subject to:~} \vmean \in \Vmean \nonumber
\end{gather}

However, in general, the objective function that drives culture metabolism is unknown.
Indeed, finding it may become the focus of study itself~\cite{schuetzSystematicEvaluationObjective2007}.
In more complex scenarios, like cancer or tissues, the problem is particularly challenging and constitutes a severe limitation for the application of $\FBA$ models.
This is aggravated because those complex scenarios are in fact the norm in nature, while optimal growth conditions are the exception.
To make matters more complicated, it may well be the case that the evolution of metabolism leads to an overall robustness across many conditions rather than a single condition-specific objective~\cite{feistBiomassObjectiveFunction2010}.

\subsection{Maximum Entropy Principle and Metabolism}
\label{sec:MaxEnt}

As mentioned before, $\FBA$ models extract from the available information (the constraints defining the metabolic spaces) a candidate flux configuration based on a given objective function.
However, an important limitation is that they provide little insights about other features connected with the culture heterogeneity.
Additionally, the $\FBA$ solution is only affected by constraints that are directly involved in defining the optimal objective value.
The remaining constraints are irrelevant, and the information encoded into them is not used.
	
A more general framework, able to take into account all the possible constraints and to provide a deeper picture of the metabolic state of the culture comes through the Maximum Entropy Principle ($\ME$)~\cite{jaynesInformationTheoryStatistical1957}.
This framework has been recently used to model the phenotypic distribution of cells in culture for several growth conditions and cultivation regimes~\cite{demartinoGrowthEntropyBacterial2016, demartinoStatisticalMechanicsMetabolic2018}.
In the context of our constraint-based model, the Maximum Entropy Principle may be formulated as follows:
\begin{gather} 
	\text{maximize arg}_{\PX} ~
		\bigg\lbrack
			-\int_{\Vcell} \PX(\vcell) log(\PX(\vcell))d\vcell 
		~\bigg\rbrack 
		\label{eq:ME} \\ 
		\text{subject to:} \nonumber \\ 
			\vcell \in \Vcell \nonumber \\ 
			\vmean = \int_{\Vcell}\vcell \PX(\vcell) d\vcell \in \Vmean \nonumber 
\end{gather}
which means that from all the feasible distributions $\PX$ we must find a distribution, which we call $\PME$, that maximizes the entropy subject to specific constraints.
Following Jayne's~\cite{jaynesInformationTheoryStatistical1957} interpretation of the principle, $\ME$ is the least biased distributions encoding all the information that we have about the system.
In other words, in absence of a mechanistic model, we consider $\PME$ to be our best guest of the real $\PX$, given the available information.

If $\Vcell$ is bounded and the constraints applied over $\Vmean$ have the simple forms $\vmean_i\le a_i$ or $\vmean_i=a_i$, where $\avec \in \Rn$ is a constant vector (such as constraints~\myref{eq:Ch_ui_stst_ineq_constraints} and \myref{eq:Ch_mu_stst_constraints}), it can be proved that $\PME$ belongs to the exponential family~\cite{demartinoStatisticalMechanicsMetabolic2018,jaynesProbabilityTheoryLogic2003a}:
\begin{equation}
	\PME(\vcell) \propto e^{\betavec^T \vcell}
	\label{eq:PME}
\end{equation}
where $\betavec \in \Rn$ is a vector ($\betavec^T$ is its transpose) of Lagrange multipliers, where each $\beta_j$ is associated with the $j^{th}$ reaction, used to select the appropriate $\PME$.
See \Appendix{sec:ME_Alg_app} for a more formal discussion.

An important point to notice, is that the $\ME$ probabilistic description of the culture metabolism allows, just like in $\FBA$, the inference of a representative flux configuration $\vmean$.
More precisely, having found $\PX(\vcell)$ from \myref{eq:ME}, we can compute the predicted average values as $\vmean = \int_{\Vcell}\vcell \PX(\vcell) d\vcell$.
However, although both methods use the same input data (the metabolic spaces), $\ME$ can be additionally queried about other metabolic features like $\eg$ cell-to-cell growth variability, flux correlations, information variation ($\eg$ due to regulation), \etc~\cite{demartinoStatisticalMechanicsMetabolic2018, tourignyDynamicMetabolicResource2020}.
This is a major advantage of $\ME$ over $\FBA$, the former exploits better the information available in the different spaces.

In part, this is possible because we can effectively decouple the constraints defining the different spaces.
Constraints at the single cell level are used in the definition of $\Vcell$ and constitute the support of the distributions in $\PX$.
Constraints at the population level (which define $\Vmean$) reduce the set of feasible distributions $\PX$, from which the one that maximizes the entropy is selected.
It is also important to remark that the $\ME$ methodology is not limited to the codification of constraints over flux averages.
Other types of population constraints can be included ($\eg$ constraints over flux variances).
A review of the utilization of $\ME$ as a general inference technique in biological problems can be found in~\cite{demartinoIntroductionMaximumEntropy2018}.

\section{Results}
\label{sec:Results}

This section contains the main results of our work.
Here we explore the advantages of using $\ME$ methods with respect to $\FBA$ in continuous cultures.
The section is divided into three subsections.
We first present a minimalist model where we explicitly discuss the difference between $\Vcell$ and $\Vmean$ and the impact of its definitions on the $\ME$ solution.
Then, we introduce a toy model of the metabolism and connect it with the dynamics of a chemostat considering a heterogeneous culture.
The numerical data about the macroscopic quantities obtained with this model will be used to feed $\FBA$ and $\ME$.
In this controlled scenario we will compare the outputs of these two approaches to clarify the differences between both of them.
Finally, we will make a similar comparison with data obtained from real continuous cultures experiments for $\Ecoli$.
With this, we test the feasibility of using $\ME$ for Genome Scale Metabolic Networks and further support the analysis done in the toy model.

\subsection{The minimum picture}
\label{sec:MEP_formulation}

\begin{figure}
	\centering
	\includegraphics[scale = 0.18]{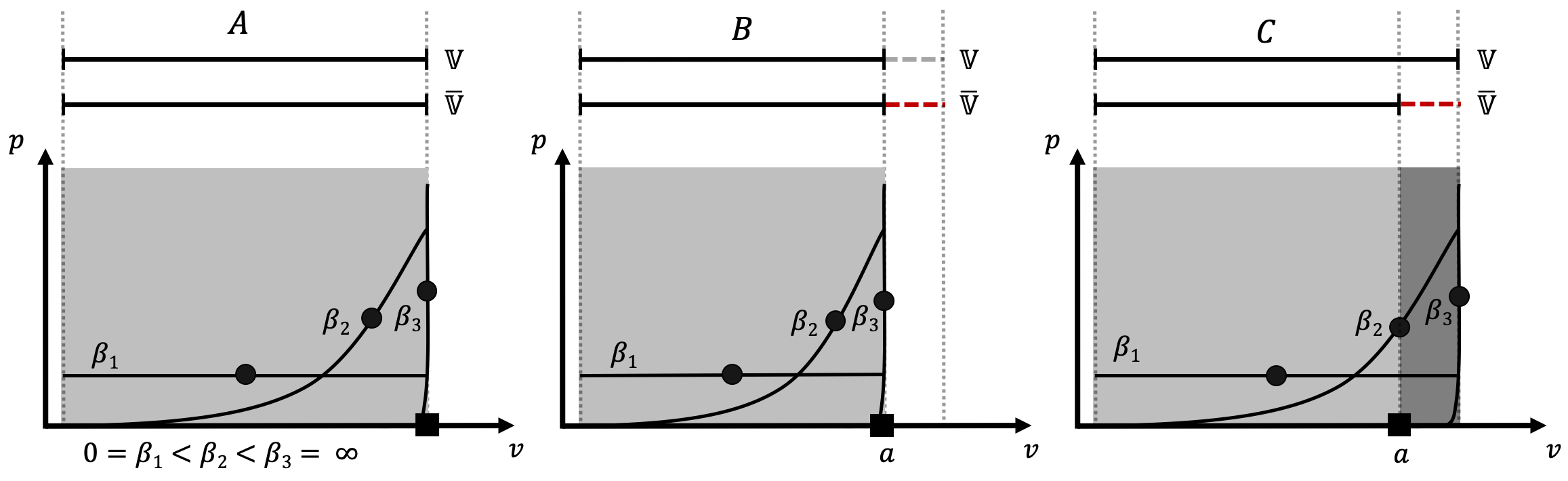}
	\caption{%
		Schemes of three different formulations of a one dimensional flux space toy model.
		In each panel, the segments (solid black lines) at the top, represent the definitions of $\Vcell \in \R$ and $\Vmean \in \R$ respectively.
		Those spaces are projected into the $x$-axis of the	graphs (following the dotted lines).
		Each graph contains three $\PME \propto e^{\beta v}$ distributions, mapped over $\Vcell$, labeled by its $\beta \in \R$ values: $\beta_1$ = 0, $\beta_2 \in (0, +\infty)$ and $\beta_3 \rightarrow +\infty$.
		The solid black circles mark the mean of the distributions (only its $x$-axis coordinates have meaning) and the solid square mark the value of an optimum $\FBA$ solution.
		Panel A show a formulation where $\Vmean = \Vcell$, whereas Panels B and C show two different formulations for encoding a real constraint over $\Vmean$, $\vmean \le \ascalar$ (dashed lines in the segments).
	}
	\label{fig:1D_flux_space}
\end{figure}

In order to gain insights about a few basic aspects of the $\ME$ formulation, we will use a model with a single free reaction, $v \in \R$ (see \Figure{fig:1D_flux_space}), which is affected by only one constraint ($lb_v \le v \le ub_v$).
This greatly simplifies the problem to which one is usually exposed because the encoding of some equations (such as \myref{eq:m_i_balance} and \myref{eq:cost}) is not necessary.

For this model, the distribution~\myref{eq:PME} has the exact form:
\begin{equation}
	\PME(\vscalar) = e^{\beta \vscalar} \psi(\vscalar)/Z \nonumber
\end{equation}
where $Z = \int \PME(\vscalar)d\vscalar$ is the normalization constant, $\beta$ is a scalar, and $\psi(\vscalar)$ is an indicator function that returns one when $\vscalar \in \Vcell$ and zero otherwise.
 
In \Figure{fig:1D_flux_space} we show a schematic representation of three different formulation of $\Vcell$ and $\Vmean$ for this uni-dimensional model.
Each panel contains a graph with the characteristic $\PME$ distributions for three distinctive $\beta$ values.
When $\beta = \beta_1 = 0$, $\PME$ is the homogeneous distribution over~$\Vcell$.
Essentially the exponential plays no role at all and each flux value $\vscalar$ is equally likely.
This is the regime of the largest entropy and maximum heterogeneity.
On the other hand, at $\beta = \beta_3 \rightarrow +\infty$, $\PME$ becomes a Dirac's delta which concentrates all the biomass density at the upper extreme of $\Vcell$.
An analogous situation is found, but at the lower extreme, if the sign of $\beta_3$ is inverted.
A Dirac's delta has the lowest possible entropy and the system is fully determined.
Finally, for $\beta = \beta_2 \in (0, +\infty)$, any intermediate  average flux value can be achieved by finding the appropriate $\beta$ value.

With these concepts clear, we can now start to dissect the differences between the three panels and their implications in the inference results for $\ME$ and $\FBA$.

First let's look at panel A.
There we do not impose any extra constraints over the fluxes and so ~$\Vmean \equiv \Vcell$.
This resembles the common $\CBM$ formulations for describing exponential growth phase of batch cultures in rich mediums, where the culture (once defined which metabolites are available) is restricted only by the intrinsic capabilities of the cell metabolism (single-cell level constraints)~\cite{varmaStoichiometricFluxBalance1994}.
In this scenario, each of the $\PME$ distributions (one for each $\beta$ value) are feasible, because all its mean values (black circles) fall inside the feasible space (shadow area).
If we consider a $\FBA$ formulation which maximize $\vmeanscalar$ over $\Vmean$, its solution~(black square) is recovers by $\ME$ at $\beta = \beta_3 \rightarrow +\infty$.
This distribution is a Dirac's delta, so our model is describing the culture as a homogeneous system (\ie~all cells display the same metabolic state).
It is important to remark that at $\beta_i\rightarrow\pm\infty$ $\ME$ will always find a mean flux vector that optimizes the flux $i$ in $\Vcell$, so it can be viewed as a generalization of $\FBA$~\cite{demartinoGrowthEntropyBacterial2016}.

The situation becomes more subtle when we introduce further restrictions (\eg~provided by data obtained from measurements in the culture).
Consider for example the new bound $\vmeanscalar \le \ascalar$ (where $\ascalar$ is a constant, resembling equation~\myref{eq:Ch_ui_stst_ineq_constraints}) that is supposed to be only applicable over $\Vmean$.
Panels B and C in \Figure{fig:1D_flux_space} show two alternative ways to introduce such constraints and their consequences.
A first approach could be the direct modification of $\Vcell$ (gray dashed line in Panel B).
In this case, although no distinction between the spaces are made ($\Vmean \equiv \Vcell$), $\Vmean$ is reduced accordingly.
However, to affect $\Vcell$ this way is unjustified.
This reduction does not follow from the rationality that leads to introduce the constraint ($\ie~\vmeanscalar \le \ascalar~\notimplies \vcellscalar \le \ascalar$).
On the other hands, Panel C shows an alternative scenario where $\Vcell$ is unaffected.
There, thanks to $\ME$'s ability to decouple the two different spaces, we can enforce the new constraint by restricting the $\beta$ values to the ones that define a $\PME$ distribution with a valid momentum.

The comparison of these two panels illustrates the consequences of adding an unjustified assumption into the space's formulation.
Although $\Vmean$ is the same in both cases, and the average flux value reported by $\FBA$ and $\ME$ solutions are not affected, other features of the $\ME$ solution do differ.
For instance, in panel B, the $\FBA$'s solution is reached at $\beta = \beta_3 \rightarrow +\infty$, whereas in the right panel, $\PME$ achieved the same mean at a lower $\beta$ value, $\beta = \beta_2$.
Therefore, the $\PME$ at these $\beta$ values are quite different.
While in the case of panel B the system is fully determined, the more rigorous spaces definition in panel C show that it is impossible to completely determine the system with the available constraints.
In the particular scenario of a nutrient-limited chemostat culture at steady state, it is not difficult to incur into such biased $\ME$ formulations.
As we already discussed in \Section{sec:nutrient_limited_cultures}, equations~\myref{eq:Ch_ui_stst_ineq_constraints} and~\myref{eq:Ch_mu_stst_constraints} imposes strong constraints over $\Vmean$, generally leading to a situation where $\Vmean \subset \Vcell$.

However, in the literature, it is not usual to find $\ME$ formulations that do make the explicit distinction between both spaces.
For instance, in~\cite{rivas-astrozaMetabolicFluxConfiguration2020}, the growth rate and the glucose uptake are directly encoded into $\Vcell$.
Both reactions upper and lower bound in equations~\myref{eq:v_j_bounds} are set to be equal to the reported experimental measurement.
Another example can be found at~\cite{fernandez-de-cossio-diazMaximumEntropyPopulation2019}.
There, a similar chemostat model is used, but only the observable growth rate ($\zmean = D$) is encoded exactly.
The model uses a single scalar $\beta$ parameter and the constraints over the exchanges are enforced by restricting $\Vcell$ directly.
A further simplification $\uicell \le \ci D/X$ is made (note that the originally derived from the dynamic model is $\uimean \le \ci D/X$), which might lead to a situation analogous to the one represented in the Panel B of \Figure{fig:1D_flux_space}.
In \Appendix{sec:Bias_app} we discuss some consequences of such simplifications.

With this understanding, we reformulate the $\ME$ model at~\cite{fernandez-de-cossio-diazMaximumEntropyPopulation2019}.
We respect the original form of the constraints over $\Vmean$ for the observable growth rate, but we also add a similar constraint over the uptake of the limiting nutrient.
The formulation will have two Lagrange multipliers $\beta$'s (i.e., two non-zero components in the $\betavec$ vector of equation~\myref{eq:PME}).
One to enforce the biomass constraint ($\zmean = D$) and the other to enforce the glucose uptake constraint ($\ugmean \le \cgDX$), see \Appendix{sec:ME_Alg_app}.
Given that most continuous cultures are nutrient-limited, this has the advantages of potentially avoiding all biases related with the uptakes by adding only an	extra parameter compared with formulation at~\cite{fernandez-de-cossio-diazMaximumEntropyPopulation2019}.

\subsection{A simple metabolic network in a chemostat}
\label{sec:CH_dyn_sim}

In order to gain further insight on the effects of the chemostat dynamics on the form of the metabolic spaces on a controlled system, we introduce a simple model to mimic the metabolism of the cell.
In this model the cell metabolism is reduced to a small size network (see Appendix \myref{sec:toy_dyn_app}) resembling the core (fermentation/respiratory/phentose phosphate) metabolic pathways.
The model has three degree of freedom, which we choose to call $\zcell$, representing the growth rate of the cell, $\ugcell$ the uptake of a nutrient (glucose), and $\uocell$ the uptake of oxygen.

To consider the presence of heterogeneity we reformulate equation~\myref{eq:CH-dX_dt}.
In the new formulation we account for the time evolution of the biomass associated with each feasible flux configuration $\vcell$, and introduce a source of heterogeneity $\epsilon \in [0,1]$.
The latter, defines a stochastic biomass redistribution over $\Vcell$.

The final, non-discretized version, of the dynamic equations for the chemostat are: 
\begin{gather}
	\frac{dX(\zcell, \ugcell)}{dt} =
	(1 - \epsilon) \zcell X(\zcell, \ugcell)
	+ \frac{\epsilon}{\Vcellvol} 
		\iint\limits_{(\zcellprime, \ugcellprime) \in \Vcell} \zcellprime X(\zcellprime, \ugcellprime) d\zcellprime d\ugcellprime 
	- D X(\zcell, \ugcell) 
	\label{eq:cont_dyn_dX_dt} \\
	\frac{d\sg}{dt} =
		- \iint\limits_{(\zcell, \ugcell) \in \Vcell} \ugcell X(\zcell, \ugcell) d\zcell d\ugcell  + (c_g - s_g)D
	\label{eq:cont_dyn_dsg_dt}
\end{gather}
where $X(\zcell, \ugcell)$ is the biomass concentration associated with the given flux configuration and $\Vcellvol = \iint_{\Vcell} d\zcell d\ugcell$ is the volume of $\Vcell$.
Equations for the evolution of the rest of external metabolites can be stated analogous to~\myref{eq:cont_dyn_dsg_dt}, but we will focus our attention only in the limiting nutrient.
For computational purposes, the metabolic space $\Vcell$ was discretized (details at \Appendix{sec:toy_dyn_app}).
	 
Equation~\myref{eq:cont_dyn_dX_dt} expresses that the rate of change of the biomass concentration associated with a given flux configuration will vary depending on the balance between the cellular growth (first two terms) and the extraction of biomass due to the chemostat dilution (last term).
The two growth terms differ in that the first is related with the growth potential associated with the given flux configuration (local) while the second depends on the growth capacity of the whole culture (global).
The global term is just the average of all local terms scaled by~$\epsilon$.
At any particular time, it contributes equally to the growth of the biomass associated with each flux configuration.
The diffusion parameter $\epsilon$ is used to control how much of $X(\zcell, \ugcell)$ growth is due to its local capacity or because of the relocation of biomass from the rest of the culture.
In the extreme $\epsilon = 1$, all flux configurations have the same growth potential irrespective of its own $\zcell$ value, which leads, if feasible, to the larger heterogeneity of the system.
In the opposite case, when $\epsilon = 0$, $X(\zcell, \ugcell)$ evolves depending exclusively on its local growth potential ($\zcell$), and no biomass reallocation is introduced.

The track of the biomass associated with each flux configuration allows the computation of the biomass distribution $\PX$ at every time step of the simulation, by defining $\PX$~as:
\begin{equation}
	\PX(\zcell, \ugcell) = X(\zcell, \ugcell)/ X 
	\label{eq:dyn_PX}
\end{equation}
where $X=\iint_{\Vcell} X(\zcell, \ugcell) d\zcell d\ugcell$ is the total biomass concentration of the culture at a given time.

Notice also that any constraint acting upon an observable restricts the set of feasible biomass distributions and influences all other observables.
For instance, equation~\myref{eq:cont_dyn_dsg_dt} is affected by the implicit constraint $\sg \ge 0$ and since, $D$, $\cg$ and $\Vcell$ are time independent, such constraint can only be implemented by dynamically transforming $\PX$ to guarantee that $\ugmean \le \cgDX$ when $\sg \rightarrow 0$.

Then, to enforces the constraints over~$\Vmean$, we must add and explicit transformation over $\PX$ that will link both \myref{eq:cont_dyn_dX_dt} and~\myref{eq:cont_dyn_dsg_dt} together.

The transformation that we use is explained in detail in \Appendix{sec:toy_dyn_app}, but in practice it reduces to: at any instant of time in which $\sg = 0$ and $\ugmean > \cgDX$ we force the equality $\ugmean = \cgDX$ to be true by re-scaling $\PX$.
With this, we link both \myref{eq:cont_dyn_dX_dt} and~\myref{eq:cont_dyn_dsg_dt} together, keeping $X$ bounded while the culture reaches non-trivial steady states in all the feasible region of the model.
It is important to remark that to propose a realistic transformation of $\PX$ is out of the scope of this work.
The only requirement for selecting the applied one was that the system, at steady state, were subject only to the defined constraints, avoiding unjustified over-restriction on~$\Vcell$ or ~$\Vmean$.

\begin{figure}
	\centering
	\includegraphics[scale = 0.13]{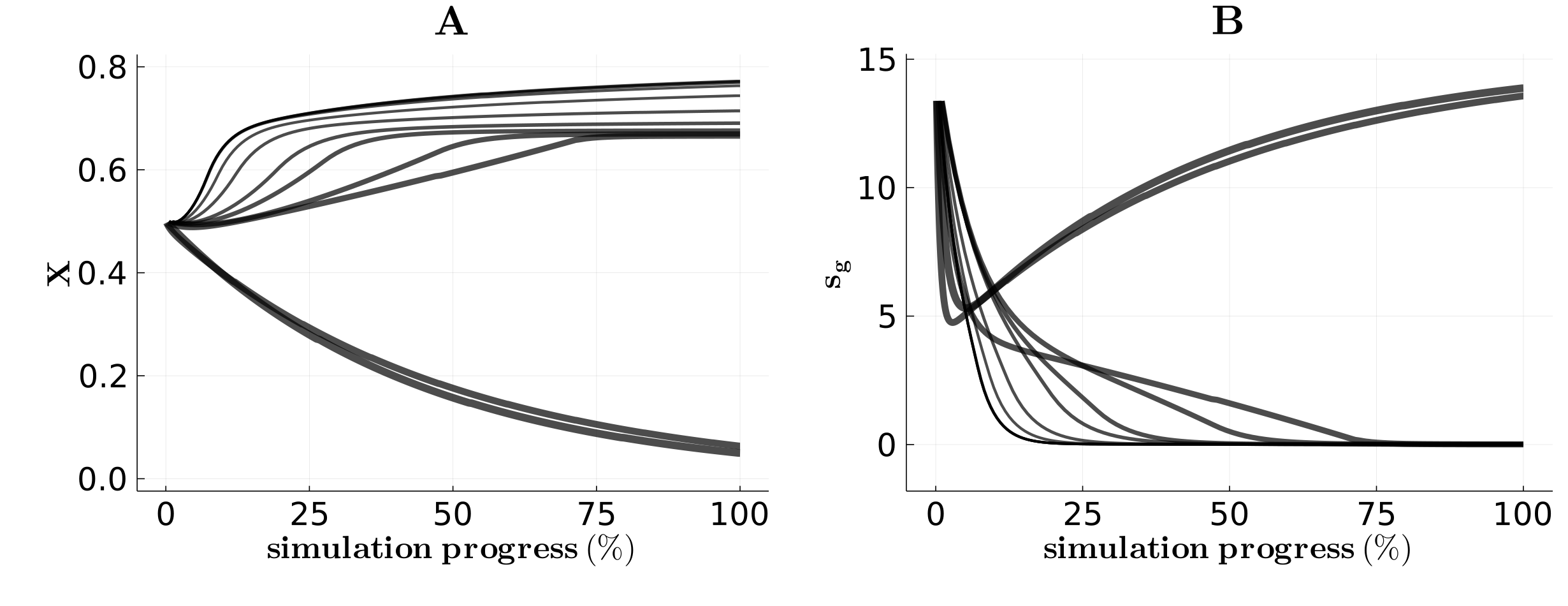}
	\caption{
		Panels A and B show the time series from the dynamic simulations of the total cell concentration ($X$) and the nutrient concentration in the vessel ($\sg$) respectively for a given $D$ and different $\epsilon$ values.
		The width of the lines are proportional to $\epsilon \in [0.001, 1]$.
	}
	\label{fig:toy_net_and_time_serie}
\end{figure}

Using this dynamics for the chemostat we performed extensive simulations of equations \myref{eq:cont_dyn_dX_dt} and \myref{eq:cont_dyn_dsg_dt} for different values of $D$ and $\epsilon$ keeping a constant $\cg$, and initial $X$, $\sg$ and $\PX$ (a uniform distribution over $\Vcell$).
As an example of the output, Panels A and B of \Figure{fig:toy_net_and_time_serie} show the time series for the total cell and glucose concentration (the limiting nutrient) in the chemostat, obtained at a constant $D$ for different $\epsilon$'s.
When the simulations reach a non-trivial steady state we computed the flux distributions and the observables needed to characterize both, the metabolism of the culture and the chemostat environment.

\begin{figure}
	\centering
	\includegraphics[scale = 0.12]{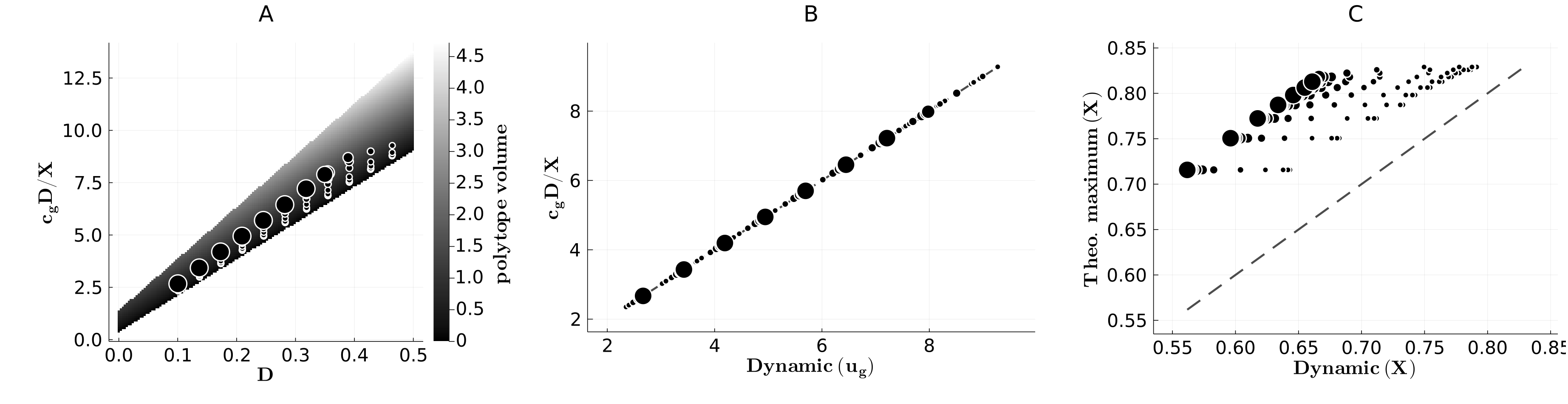}
	\caption{
		The Panel A shows a heat map measuring the volume of $\Vmean$ (length of the vertical short solid line in \Figure{fig:pol_stst_chemostat} Panel B) as a function of the steady state parameters $D$ and $\ugmean$ upper bound.
		Darker regions mean smaller volumes (log scale).
		Over the map, the locations of the steady state parameters of a set of simulations are represented by markers.
		Panel B shows a correlation between the dynamic $\ugmean$ at steady state and the glucose uptake value $\cgDX$.
		Panel C shows a correlation between the dynamic $X$ at steady state and the theoretical maximal $X$ given the constraints bound values.
		The size of the markers are proportional to $\epsilon \in [0.001,1]$.
	}
	\label{fig:toy_space_config}
\end{figure}

In \Figure{fig:toy_space_config}, Panel A, we show a heat map that represents the volume of $\Vmean$ in the ($\zmean$,~$\ugmean$) subspace as a function of the constraint bounds $D$ and $\cg~D/X$.
Darker areas in the map mean that the culture steady state configuration is near to the maximal restrictive power of the environmental constraints (and so the minimal $\Vmean$ volume).
The markers in the figure represent the location of the chemostat parameters of a set of the simulations at steady state.
The size of the markers are proportional to $\epsilon \in [0.001,1]$.

As can be appreciated, for a particular $D$ value, the larger the $\epsilon$ used in the simulation the larger the volume of $\Vmean$ at steady state.
This can be explained by the combination of two factors: I)~The tendency of the culture to reach the steady state when the limiting-nutrient is depleted, 
and II)~the redistribution of biomass over $\Vcell$ due to heterogeneity.
The first factor can be explained by the feasible dynamic's tendency to increment $X$ indefinitely in a nutrient-unlimited condition (see~\Appendix{sec:unlimited_dyn_app}).
The culture will only stop growing (and so the steady state reached) when the limiting nutrient is depleted.
For the simulations, this implies that at steady state, the culture will be consuming glucose at a rate close to the upper limit ($\cgDX$).
Panel B of \Figure{fig:toy_space_config} shows a correlation that directly supports this claim.

The second factor can be explained if we study the relation between $X$ and $\ugmean$ at steady state.
From the uptake constraint ($\ugmean \le \cgDX$) we can see that at a given glucose feed rate ($\cg D$), $X$ reach a maximum when $\ugmean$ is minimal.
This is equivalent to say that $X$ will be maximal when all cells are consuming glucose at its maximum feasible $\zmean$/$\ugmean$ yield (dashed line in \Figure{fig:pol_stst_chemostat}).
Such necessary homogeneity directly links the culture's heterogeneity with $X$ at steady state.
If the stochastic redistribution of biomass is not null ($\epsilon > 0$), metabolic states with lower yields will be occupied by the cells (see global term in \Equation{eq:cont_dyn_dX_dt}).
In those cases the culture will still tend to maximize $X$, but the heterogeneity will prevent it to reach the optimum value (and so the minimum $\Vmean$ volume) at steady state.
Panel C of \Figure{fig:toy_space_config} show such tendency by correlating the results from the simulations with the theoretical maximum $X$.
This is estimated computing the minimum $\ugcell$ value compatible with the given growth rate $\zcell = D$ and using the glucose-limited uptake bound ($max(X) = \cg D / min(\ugcell)$.

Given those results, if we revisit \Figure{fig:pol_stst_chemostat}, all glucose-limited steady states will be located inside $\Vmean$ at the circle markers ($\ugmean \approx \cgDX$).
Additionally, a culture with minimal heterogeneity ($\epsilon \rightarrow 0$) will display a $\Vmean$ configuration at steady state as represented in Panel A.
Any other ($\zmean$,~$\ugmean$) pair is disallowed due to the $\zmean = D$ constraints and the maximization of $X$.
If significant heterogeneity is introduced ($\epsilon \gg 0$), the steady state will be configured as represented in Panel B.
Note that the culture ($\zmean$,~$\ugmean$) will be far from the optimum $\zmean$/$\ugmean$ yield (mark as a solid triangle).
It is important to remark that no constraints are been formulated to control the distribution of biomass into other free dimensions.

We now use the results of these simulations to test our inference methods.
Together with $\ME$ we present results using five formulations of $\FBA$.
The first $FBA$'s objective function we use is the common maximization of the biomass.
Here, we do not force the constraint over the growth mean~($\zmean = D$).
The other four $\FBA$ stack of objective functions account for each one of $\Vmean$'s vertices.
Because of the simplicity of the toy model and the chemostat constraints at steady state, $\Vmean$ has only four vertices, and so, $\FBA$ (as formulated in~\myref{eq:FBA_formulation}) will yield only four possible solutions (one for each vertex~\cite{orthWhatFluxBalance2010}).

\begin{figure}
	\centering
	\includegraphics[scale = 0.095]{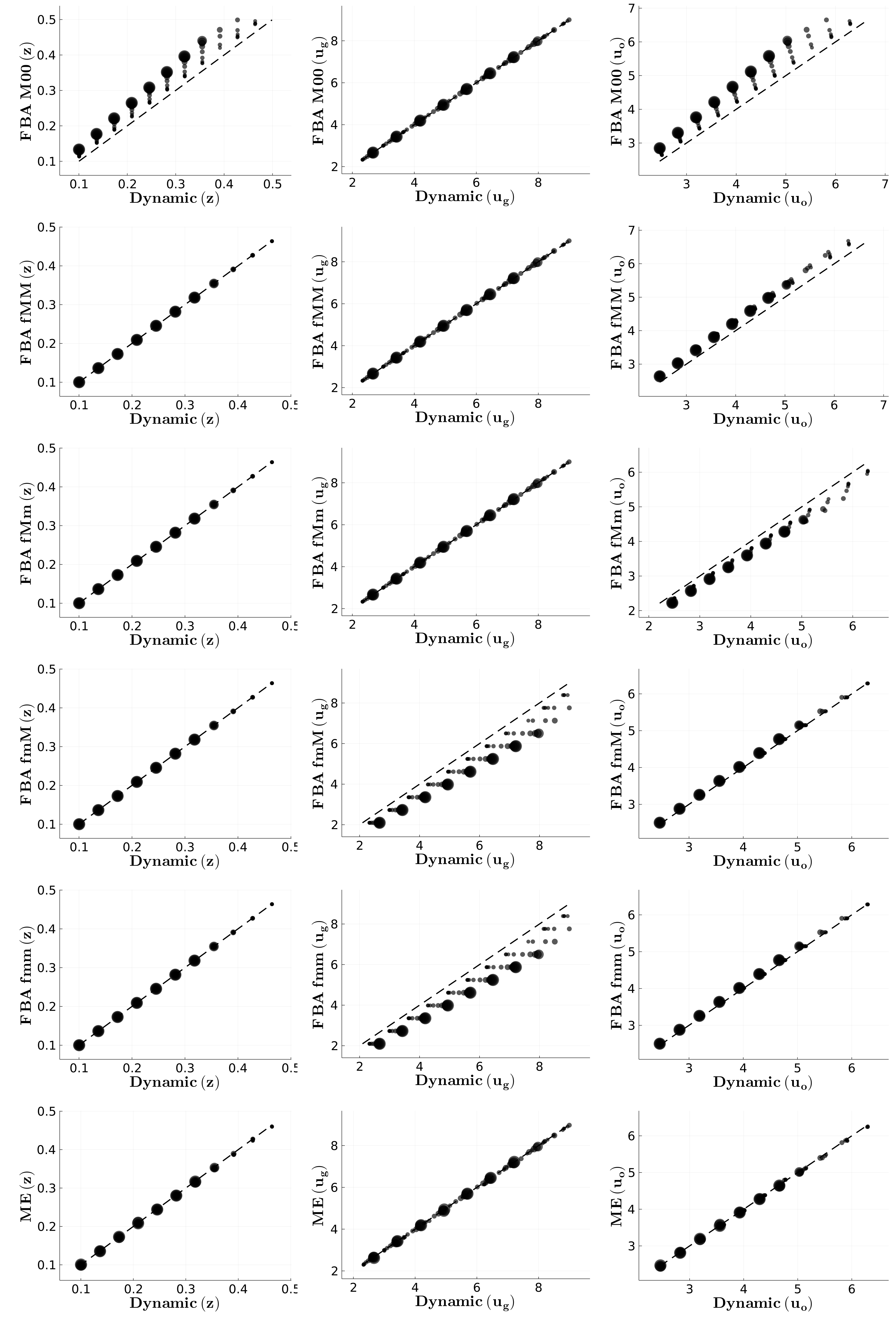}
	\caption{
		Correlations between the dynamic (x-axis) and inferred mean values (y-axis) for the free fluxes $\zcell$, $\ugcell$ and $\uocell$ of the toy network.
		Each row shows the results of one inference method and each column of one free flux.
		In the case of the $\FBA$ formulations, we specify the sequence of objective functions required to determine a solution
		by using a character triple where: 'm' means minimization, 'M' maximization, 'f' that the flux was fixed to a given value and '0'
		that no further action was required.
		The position of the character express the action over $\zcell$, $\ugcell$ or $\uocell$ respectively
		(Ex: 'M00' means that the maximization of $\zcell$ lead to a single solution).
		The size of the markers encode the value of $\epsilon \in [0.001,1]$.
	}
	\label{fig:toy_corrs}
\end{figure}

\Figure{fig:toy_corrs} shows the correlations between the artificial data and the models for all simulations that reached a non-trivial steady state condition.
The mean value for each free flux ($\zcell,~\ugcell,~\uocell$) is computed from the dynamic biomass distribution~\myref{eq:dyn_PX} at steady state, 
the inferred $\ME$ distribution \myref{eq:ME} and the solution of the $FBA$ optimization~\myref{eq:FBA_formulation}.
In this case, each $\PME$ is inferred by finding the two beta parameters that made the distribution fulfill both observable constraints ($\zmean = D$ and $\ugmean \le \cgDX$) and maximizes the entropy (see \Appendix{sec:ME_Alg_app}).
Each row correspond with a different inference technique and each column with a free flux.

The first row of the figure presents the results for the $\FBA$ formulation $\FBA^{(M00)}$ (see \Figure{fig:toy_corrs} caption for notation details), 
which maximizes the biomass rate (it does not include the $\zmean = D$ constraint).
This formulation consistently overestimates $z$, but correctly predicts the glucose uptake~$\ugcell$.
The maximization of $\ugmean$ and the additional overestimation of $\uocell$ in $\FBA^{(M00)}$ is consistent with the structure of the network and the maximization of $z$.
For instance, the consumption rate of glucose and oxygen are proportional to the $ATP$ production rate, that is a reactant in the biomass equation.
In \Figure{fig:pol_stst_chemostat} Panel B, this solution is located in the squared marker.
In our model of the chemostat, for such solution to be valid (and the objective function to be useful), the heterogeneity must be zero.

From the second to the fifth row of \Figure{fig:toy_corrs} the results of the rest of $\FBA$'s formulations is shown.
These formulations respect the chemostat constraint over $z$, as is trivially appreciated in the correlations of the first column.
Although the formulations which maximize $\ugmean$ ($\FBA^{(fMm)}$ and $FBA^{(fMM)}$) reproduce two of the three free fluxes of the simulations, in general $\FBA$ was incapable of capturing the whole metabolic state of the culture.
In particular, no $\FBA$ formulation was able to infer $\uomean$ correctly.
As stated before, the chemostat steady state and the glucose-limiting condition are only constraining $\Vmean$ in the ($\zmean$,~$\ugmean$) subspace.
If $\Vmean$ is degenerated in other dimensions, the observed value is not necessarily an optimum.
The error induced increases with $\epsilon$ (in \Figure{fig:toy_corrs}, the value of~$\epsilon$ is proportional to the size of the markers), \ie~stochasticity leads to heterogeneity and this influences negatively the performance of $\FBA$.
This is specially significant given that, as we mentioned, these formulations exhaust the space of possible solutions which linear objective functions can yield for this simple model.
That is, it is not the ignorance of the correct $\FBA$'s optimization function that is causing these results, it is the fact that the culture is not in an optimal metabolic state compatible with the known constraints.

In the last row of \Figure{fig:toy_corrs} we present the results of $\ME$.
The panels show that, $\ME$ is able to reproduce all the observables independently of the stochasticity of the metabolism.
Notice however, that it does so, without explicit inputs about any optimization function followed by the cell.
Such good inference results support the idea that the simulation observables were affected significantly only by the constraints used in the definition of the metabolic spaces included in $\ME$.
More importantly, it suggests that adding further assumptions will likely bias the $\ME$'s solution rather than improve it.

\begin{figure}
	\centering
	\includegraphics[scale = 0.11]{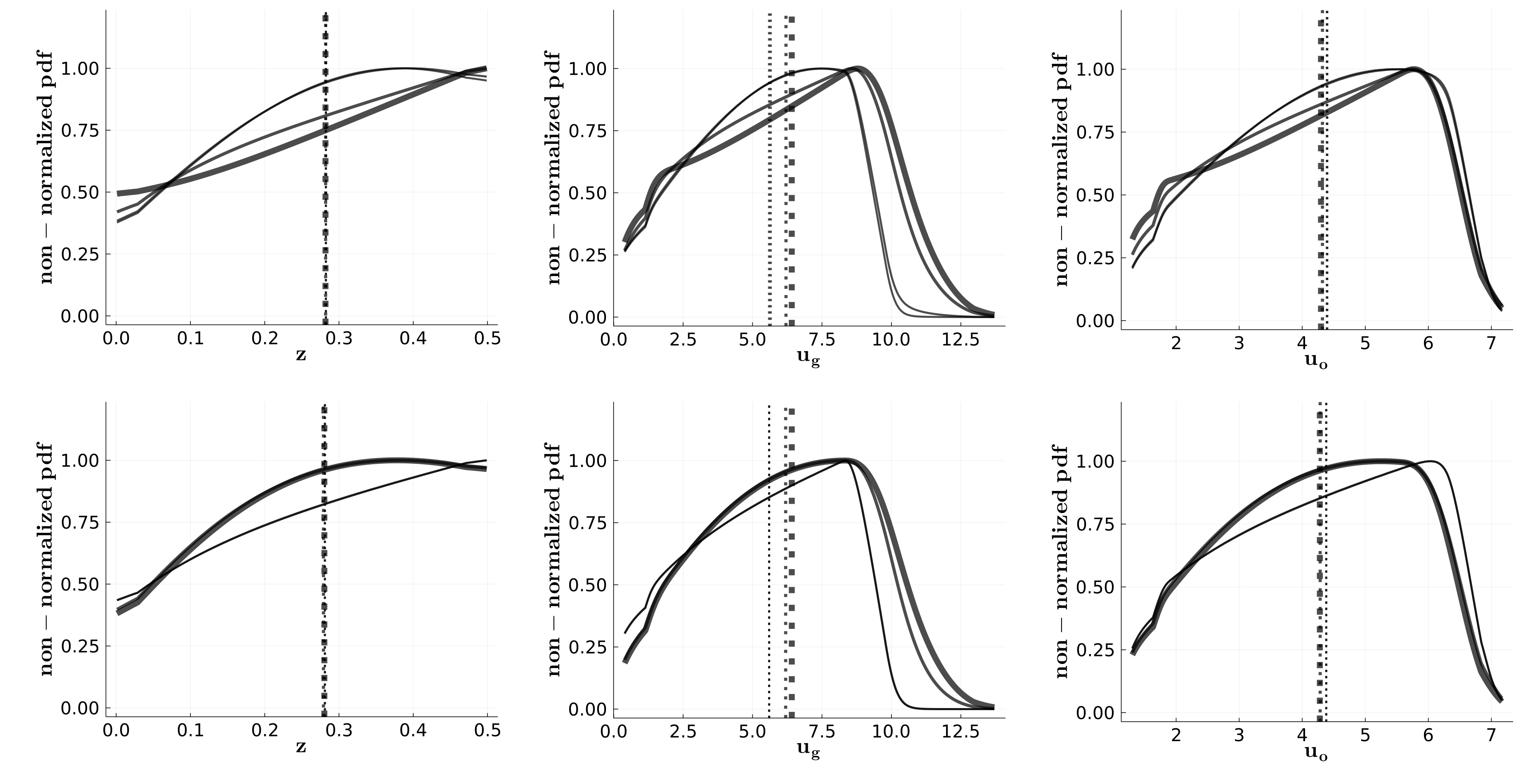}
	\caption{
		Steady state marginal distributions for $\zcell$ (left column), $\ugcell$ (center column) and $\uocell$ (right column) from the
		toy model dynamic (top row) and $\ME$ (bottom row).
		All results are at a fixed $D$ value, while distributions at different $\epsilon$ are shown.
		The dotted lines mark the mean for each distribution.
		The width of the lines are proportional to $\epsilon \in [0.001,1]$
	}
	\label{fig:toy_marginals}
\end{figure}

One of the advantages of $\ME$ over $\FBA$ is that its solution is a full probabilistic description of the metabolic state of the culture.
\Figure{fig:toy_marginals} shows the marginal distributions for the free fluxes at different values of $\epsilon$ for simulations at a fixed $D$.
The upper row shows the distributions produced by the dynamical simulation and the lower row the ones inferred using $\ME$.
As we already showed before, $\ME$ infers correctly the mean values (dotted lines) of the distribution, but although we can see that it also describes quite well the real shape of the distributions, there are differences.
This gives us an important insight: even if we encode correctly the constraints that are defining $\Vcell$ and $\Vmean$, and if these definitions really describe the boundary of the experiments (the simulations), our $\ME$ formulation does not include information about constraints acting over higher order moments of the distributions.
In this particular case, the arbitrary $\PX$ transformation introduced on the dynamic for enforcing the moment constraints, although not affecting $\Vcell$ or $\Vmean$, is generating a non-uniform effect that differentiate the distribution computed directly from the simulations from the one inferred through $\ME$, $\PME$.
If those extra constraints were encoded in the model, $\ME$ is expected to recover the distributions completely~\cite{jaynesProbabilityTheoryLogic2003a}.
	 
\subsection{Observable flux configuration inference in E. coli: FBA versus ME}

In this section we produce an analysis similar to the one made in the previous section, but using a genome scale metabolic network~\cite{reedExpandedGenomescaleModel2003} and a set of real experimental observations obtained during $\Ecoli$ glucose-limited continuous cultures \cite{kayserMetabolicFluxAnalysis2005, nanchenNonlinearDependencyIntracellular2006, folsomPhysiologicalProteomicAnalysis2014}.

\begin{figure}
	\centering
	\includegraphics[scale = 0.115]{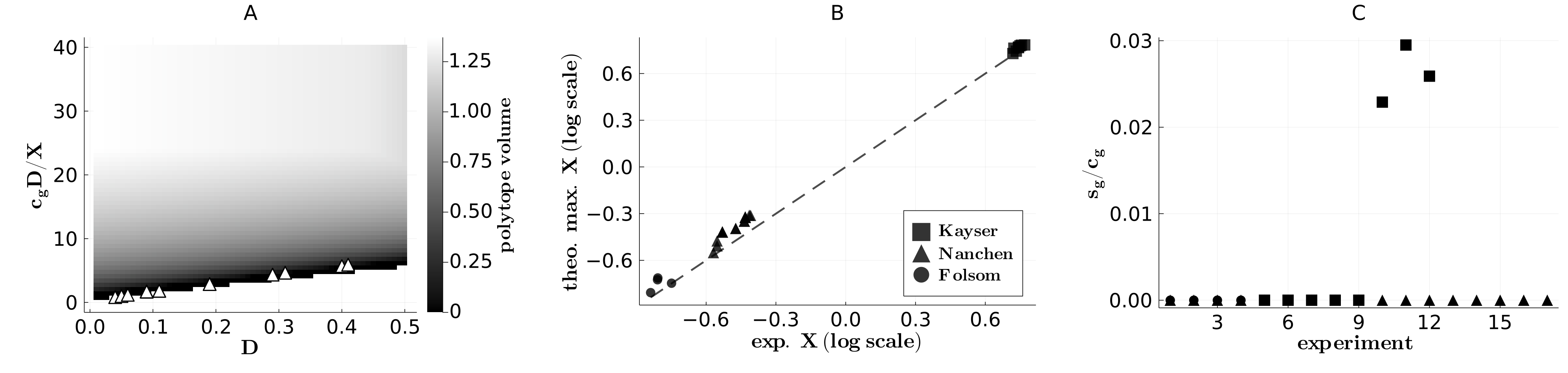}
	\caption{
		The left panel (A) shows a heat map of the polytope ($z$,~$\ug$) projection box volume (log scale) as a function of the steady state parameters, $D$ and $\cgDX$.
		The triangles show the experiment location in such space.
		Dark regions correspond with the scenario described in \Figure{fig:pol_stst_chemostat} Panel A.
		Data shown only for $\Nanchen$~\cite{nanchenNonlinearDependencyIntracellular2006}, but the rest of the data sets displayed a similar behavior.
		On the central panel (B), it is shown a correlation (log scale) of the theoretical maximum $X$ as computed using the metabolic network with respect to the experimentally reported.
		Finally, the right panel (C) shows the residual glucose at the vessel ($\sg$) relative to the concentration in the feed medium ($\cg$) for all experiments.
		Marker shape denote the experimental data source.
	}
	\label{fig:glc_limited_study}
\end{figure}

In our model, $\Vmean$ is constrained only in the ($z$,~$\ug$) subspace (see \Section{sec:nutrient_limited_cultures}).
Aiming to elucidate how much the observable space is restricted in real glucose-limited cultures, we contextualized the genome-scale metabolic network according to the experiments conditions (see \Section{sec:Ecoli_mat} for details).
Later, we compute the volume of the $\Vmean$ ($z$,~$\ug$) subspace.
\Figure{fig:glc_limited_study}, panel A, shows a heat map that illustrates such volume as a function of the parameters of the chemostat steady state (analogous to the one in \Figure{fig:toy_space_config}).
The area outside the heat map is unobservable based on all defined constraints.
As can be noticed, all experiments (triangular markers) are very close to the limit of feasible space (darker region).
In this case, the set of steady state parameters approaches the restrictive limit imposed by the constraints, similarly to the scenario described in panel A of \Figure{fig:pol_stst_chemostat}.
The heat map is only showing the results for one data set, but the others displayed a similar behavior.

Although, in principle, the experiments can be located in any point within the feasible space (shadow area) they all sit at the border of the feasible/unfeasible transition.
As mentioned before, this transition coincides with the maximum theoretical value expected for $X$.
Panel B of the figure supports this idea: the cultures are close to the theoretical maximum $X$ derived from the experimental conditions and the used metabolic network.
As before, the  maximum is computed by finding the minimum $\ugcell$ value  compatible with the given growth rate $\zcell = D$ and deriving it from the glucose-limited uptake bound ($max(X) = \cg D / min(\ugcell)$).

As stated before, such maximization of $X$ is only possible if the culture is consuming glucose at a rate that nearly matches the nutrient input feed rate ($\cg D$), which implies that most of the residual glucose in the vessel is depleted.
For completeness, in Panel C of the figure, the reported concentration of residual glucose in the vessel $\sg$ relative to the feed concentration $\cg$ is shown.
As can be observed, all experiments had imperceptible or small amounts of glucose  at steady state (for the higher values it is less than 4\% of the feed concentration).

Finally, we used $\ME$ to infer a biomass distribution $\PME$ for each experimental condition.
In order to do that, and in analogy with the simple model above, we computed the two free components on the $\betavec$ vector that allow us to enforce the moment constraints ($\zmean = D$ and $\ugmean \le \cgDX$) and maximize the entropy (see Appendix ~\myref{sec:ME_Alg_app}).
Due to the large number of variables involved, in this case, the $\PME$ functional becomes intractable, and so, we use $Expectation\ Propagation$~\cite{minkaExpectationPropagationApproximate2013} in order to approximate these distributions (see \Appendix{sec:EP_app} for details).
Additionally, we used a set of four $\FBA$ formulations as reference to compare the performance of $\ME$.
Since for a genome-scale network $\Vmean$ is not as simple as in the toy model, the full set of possible lineal $\FBA$ solutions becomes intractable.
We first introduced two objective functions common in the literature.
It has been found that for chemostat cultures, the maximization of $atp$ or biomass yield (equivalent to the minimization of $\ugmean$) objectives provide better results approximating experimental data than other tested functions~\cite{schuetzSystematicEvaluationObjective2007}.
In addition to those two, we defined the maximization of the glucose uptake $\ugmean$ (motivated by the glucose-limited condition) and the traditional maximization of biomass rate as objective functions to be tested.

\begin{figure}
	\centering
	\includegraphics[scale = 0.075]{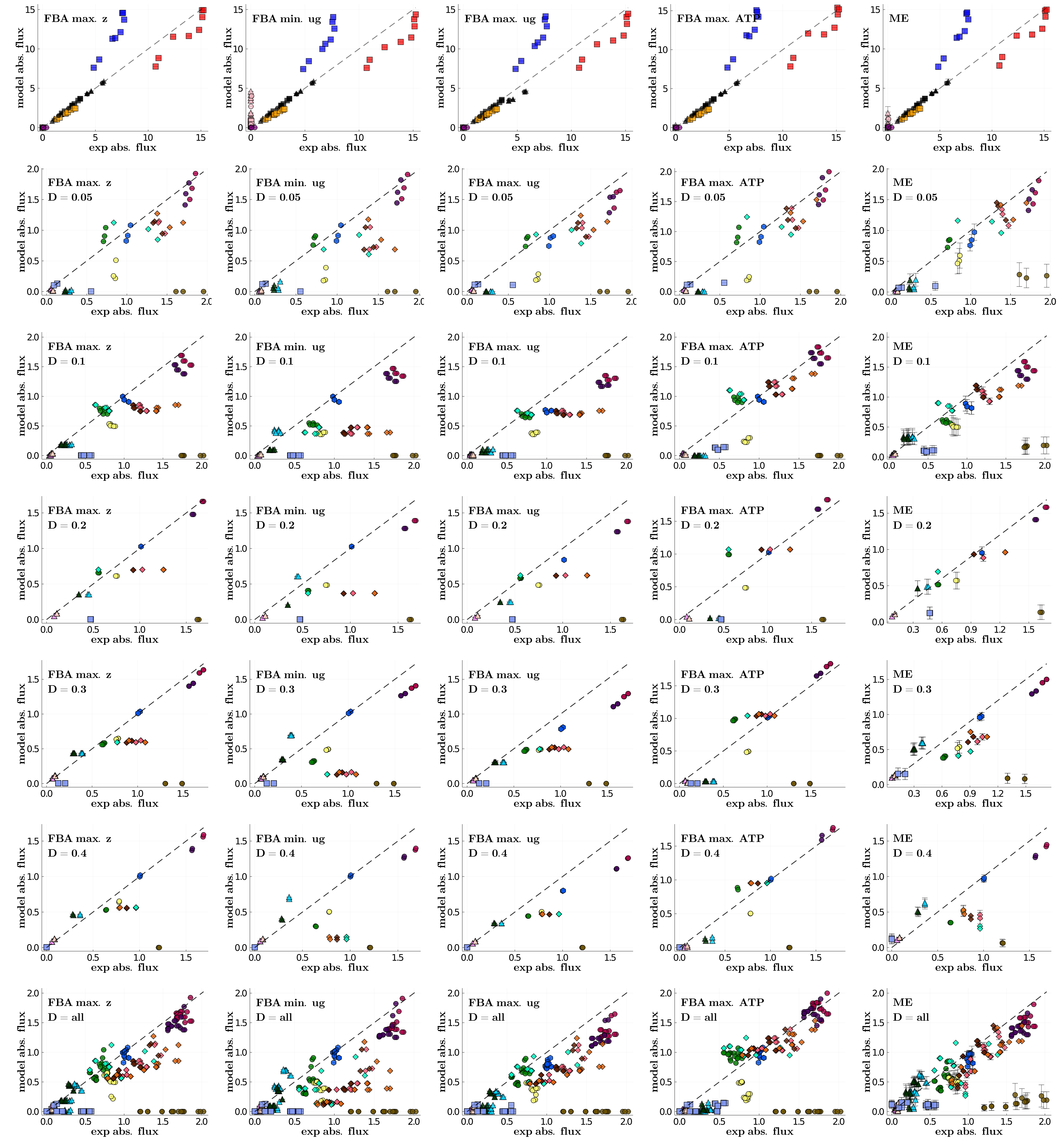}
	\caption{
		Experimental (x-axis) \emph{v.s.} model predicted fluxes (y-axis), for a set of $\FBA$ formulations and $\ME$.
		The first row shows exchange fluxes reported in~\cite{kayserMetabolicFluxAnalysis2005, nanchenNonlinearDependencyIntracellular2006, folsomPhysiologicalProteomicAnalysis2014}.
		The rest of the rows show some inner fluxes reported in~\cite{nanchenNonlinearDependencyIntracellular2006}.
		Each row corresponds to a different dilution rate.
		The last row includes all internal flux correlations.
		Different subsystems are signalized by the shape of the marker meanwhile different colors denote individual reactions.
		The legend is as follows: acetate (gray), CO2 (blue square), glyoxylate cycle (pink), pyruvate kinase $PYK$ (dark green), Krebs cycle (diamond shaped) and glycolysis (circle shaped), pentose phosphate pathway (triangle shaped).
	}
	\label{fig:corr_collage}
\end{figure}

In \Figure{fig:corr_collage}, the correlations for a set of inferred observable fluxes (exchanges and internals) with respect to the experimentally reported are shown.
The first row of the figure shows the data for the exchange fluxes (absolute values) reported in all the used data sets.
There, similar results were obtained from all the used inference techniques.
The main difference appears in the value of the acetate (gray markers) production rate.
For all the studied cultures, the dilution rate was smaller than $0.5$ $\Dunits$ (below the acetate switch~\cite{basanOverflowMetabolismEscherichia2015}) and therefore, experiments do not report acetate production.
However, the $\FBA$ model that maximizes $\ugmean$ (second column) a $\ME$ (last column) inferred non-zero acetate production rates for all data sets, although the $\ME$ ill-prediction is less pronounced.
Additionally, all methods show poor correlations of the produced CO2 (blue square markers).
A sustained overproduction of CO2 is predicted consistently by the network.
All gases exchange data come from one data set where culture's experimental carbon recovery was satisfactory ($> 93\%$), and no carbon rich byproduct, other than CO2 and biomass itself, was produced~\cite{kayserMetabolicFluxAnalysis2005}.
This suggests that such overproduction of carbon-rich byproducts can be related with an underestimation of the carbon requirements in the biomass equation, which can also affect the acetate ill-prediction mentioned before.

In the rest of the rows of the \Figure{fig:corr_collage}, we show the correlations corresponding with internal fluxes reported at~\cite{nanchenNonlinearDependencyIntracellular2006}.
Each row is related to experiments performed at different average $D$ values.
The last row corresponds with the joint correlation of all experiments.
Only two of the reported fluxes were predicted significantly wrong and consistently by all methodologies: the flux through the glyoxylate cycle (pink markers) and the pyruvate kinase $PYK$ (dark green markers).
Even though, $\ME$ always predicted a non-zero flux whereas $\FBA$ generally assigned an exact zero value through them.
The glyoxylate cycle in particular, is notoriously known to be difficult to predict by linear optimization formulations~\cite{demartinoStatisticalMechanicsMetabolic2018, rivas-astrozaMetabolicFluxConfiguration2020}, which frequently assign a zero flux to it.

In the internal correlation plots, different metabolic subsystems are labeled with different marker shapes (colors identify particular reactions).
The main insight that can be appreciated in the correlations is that there is not a single $\FBA$ formulation performing better than $\ME$ for all subsystems in all dilution rates.
Although it can be noticed in the joint correlation (last row), a careful check of each case better supports such conclusion.
For instance, the $\FBA$ formulation that maximizes the $ATP$ production (fourth column), has similar or slightly better correlations for many of the fluxes in the Krebs cycle (diamond shaped markers) and glycolysis (circle shaped markers).
But, it systematically underestimates the fluxes of the pentose phosphate pathway (triangle shaped marker).
In an opposite case, the formulation which minimizes the glucose (second column), improves the inference of the pentose phosphate pathway but ill-predict the Krebs cycle reactions.
A similar analysis can be made with the rest of the formulations.
This situation resembles the results obtained in the toy model section.
Suggesting that the metabolic state of the culture might be not well describes by a polytope vertex (an optimum).

\section{DISCUSSION}

We start the discussion by analyzing some limitations of our approach.
In the previous sections we highlighted the relevance of codifying the different constraints into its corresponding spaces.
Although we handle properly the constraints related with the chemostat dynamic, we introduced several simplifications into the definition of $\Vcell$.
We take information that is actually based on macroscopic measurements to be representative of each cell.
This hides a culture homogeneity assumption.

The most significant is related with the definition of the biomass equation, which is determined by experimentally measuring the average cellular composition~\cite{feistBiomassObjectiveFunction2010}.
In principle, this constraint (dashed line in the \Figure{fig:pol_stst_chemostat}) should be considered to be affecting only $\Vmean$.
But we made the widely adopted simplification~\cite{fernandez-de-cossio-diazCharacterizingSteadyStates2017, fernandez-de-cossio-diazCellPopulationHeterogeneity2019, demartinoGrowthEntropyBacterial2016, muntoniRelationshipFitnessHeterogeneity2021} of taking it as a hard constraint over $\Vcell$.
In the current formulation, $\ME$ is only capable to encode average constraints over the reaction bounds (like equations~\myref{eq:Ch_ui_stst_ineq_constraints} and~\myref{eq:Ch_mu_stst_constraints}), not balance constraints such that the biomass equation.
In the particular case of a limiting-nutrient chemostat culture, this might be a fundamental source of bias given the tendency of the culture to maximizes $\zmean$/$\ugmean$ (see results at \Figure{fig:glc_limited_study} Panel A).
At this point, the degeneracy of $\Vmean$ is minimal, and so, the variability lost by the biomass simplification might be significant.
We leave this question open for future studies.
A similar situation occurs in the formulation of the cost constraints in equation~\myref{eq:cost}.
In particular, the definition of each cost weight~$\ajfwd$~and~$\ajbkwd$ depends on the total observable protein mass fraction of the cells~\cite{begIntracellularCrowdingDefines2007}.
This is another balance constraint that can not be encoded into $\Vmean$ using the current $\ME$ formulation.
A deeper analysis of the consequences of the simplifications is found in~\Appendix{sec:Bias_app}.
Also, although the studied fluxes are representative of important metabolic pathways, our system is still heavily under-determined. We only have access to approximately $10^1$ experimental fluxes in a network with more than $10^3$ reactions.

Having said that, and motivated by the generalization capability shown by $\ME$, it is worthwhile to ask: When is $\ME$ actually relevant within this context? What is needed for it to be a good descriptor of the culture metabolism?
In principle, all the effective constraints that are acting upon the evaluated properties ($\eg$ the culture observable flux configuration) must be known and included correctly in the model formulation~\cite{jaynesInformationTheoryStatistical1957}.
Fortunately, in the context of metabolic models, the data necessary for formulating environmental constraints are commonly available ($\eg$ medium composition, cellular concentration, culture observables, \etc).
However, although progress in this area has been substantial, metabolic models generally lack some information needed for a complete formulation of internal constraints ($\eg$ kinetic parameters of enzymes, influence of the regulatory network, completeness of the stoichiometric network, \etc)~\cite{palssonSystemBiologyConstraintbased2015}.
The relative relevance of both types of constraints in a particular experimental condition determines the effectiveness of the inference method.

For example, in a rich medium, like a batch culture, the environmental constraints are not too strong.
The cells are growing in a context without limiting nutrients, and therefore they can potentially display a wide range of phenotypic behaviors ($\Vmean \equiv \Vcell$).
In this case, the unknown non-environmental constraints ($\eg$ regulatory or kinetic) are defining the behavior of the culture.
If this is the case, a $\ME$ formulation that lacks such decisive constraints must lead to a solution that poorly describes the observed phenotypic state of the culture.
Traditionally, this issue has been addressed by introducing further constraints based on available experimental data ($\eg$ fixing a fraction of the  fluxes~\cite{muntoniRelationshipFitnessHeterogeneity2021}) or in the case of $\FBA$ by defining an objective function (or a stack of them).
In the latter case, the objective function tries to represent those unknown non-environmental constraints that are driving the system to a specific state inside the very degenerated feasible solution space.

On the other hand, in a nutrient-limited chemostat culture at steady state, the known environmental constraints (as defined in our model) lead to a restricted observable space~($\Vmean \subset \Vcell$).
So, there is less room for unknown non-environmental constraints to significantly affect the observables.
In this case, it is natural to assume that a $\ME$ formulation that includes such information should be a good descriptor of the culture metabolic state.

Such a scenario was modeled into the chemostat simulation.
By construction, $\Vmean$ was restricted only by known environmental constraints.
More interestingly, the chemostat constraints potentially determined $\Vmean$ in only two of the free dimensions, so $\Vmean$ was generally degenerated.
This degeneracy was not due to missing information, but because of the added stochasticity resembling the natural stochastic phenomena characteristics of cellular cultures~\cite{elowitzStochasticGeneExpression2002, fernandez-de-cossio-diazCellPopulationHeterogeneity2019, huhRandomPartitioningMolecules2011, wangClonalEvolutionGlioblastoma2016, tzurCellGrowthSize2009} in the absence of further constraints.
Indeed, in this case,  $\ME$  accurately describes the steady state~$\vmean$ of the system (see results at \Figure{fig:toy_corrs}).
On the contrary, such a situation had catastrophic consequences for $\FBA$.
In practice, it literally means that no further assumption (optimization) needs to be formulated for describing the properties of the system. All the required information was already contained in the metabolic spaces formulations.

Increasing further the stochasticity in the simulation of the chemostat, we study the more general case where $\Vmean$ is degenerated even in the ($\zmean$,~$\ugmean$) subspace.
It revealed a link between the heterogeneity and the size of $\Vmean$ at steady state.
In the simulation, stochasticity affects two important features of the steady state: I) its feasibility (see \Appendix{sec:Bias_app}), and II) how large can be $X$ (see results at \Figure{fig:toy_space_config}).
As mentioned, the maximum $X$ can only be reached by a culture displaying a maximum $\zmean$/$\ugmean$ yield (for a glucose-limited case).
The stochasticity prevents that from happening, by forcing the culture to allocate biomass at sub-optimal states.

Finally, the dynamic simulation provides a possible mechanistic explanation on how the steady state constraints become so relevant.
The simulations demonstrated that the system, when it is feasible, displays a typical tendency to use the full carrying capacity of the medium  (accumulating $X$ until the limiting nutrient is depleted).
Although we provided a very simple dynamical model to test this idea, the results can be generalized to more realistic scenarios.
For example, introducing a biomass rate maximization constraint (a popular regulatory constraint for bacteria~\cite{feistBiomassObjectiveFunction2010}), the dynamics of the system will change, but the space of observables is still determined by the same environmental constraints ($\zmean = D$ and $\ugmean \le \cgDX$).
This means that unknown and complex constraints could be driving the dynamic phase of the culture, but at the steady state, its consequences over $\Vmean$ are ultimately summarized in the value of $X$.
This is because, as mentioned before, $\Vcell$, $D$ and $\cvec$ are usually considered constant during the culture, and $X$ is the only variable dependent on the dynamics that influence the chemostat constraints.
A similar picture was discussed in~\cite{fernandez-de-cossio-diazCharacterizingSteadyStates2017}, where a related dynamical model is studied.
There, the authors established that the ratio between cell concentration and dilution rate is the control parameter fixing the steady state properties of the chemostat.
The conclusion can be extended to more complex scenarios such as multi-stable regimes~\cite{fernandez-de-cossio-diazCharacterizingSteadyStates2017}.

We may extrapolate some insights gained in the dynamic simulation analysis into the interpretation of the results obtained using the realistic network and the experimental data.
A first noticeable result was the location of the experiments appear close to the maximal theoretical $X$, defined by the metabolic network and the culture conditions~(see ~\Figure{fig:glc_limited_study}).
On the simulation, the culture's heterogeneity was inversely proportional to $X$ (see~\Figure{fig:toy_space_config}).
So, its maximization in the experiments suggest that the $\Ecoli$ cultures had the minimal possible heterogeneity as result of the restrictions imposed by the chemostat constraints.
This also means that the experimentally feasible $\Vmean$ is minimum, \ie~it is the more informative state yield by the environmental constraints~\cite{jaynesProbabilityTheoryLogic2003a}.
It is sensible then to ask: Is this enough information for describing the culture observables?
As mentioned just before, a positive answer would imply that $\ME$ must be able to recover such culture property.
The correlations result at~\Figure{fig:corr_collage}, although not conclusive due to the noted limitations of our model, point into this direction.
We might be in the desirable situation where the most significant restriction are the known environmental constraints.

A possible biological interpretation is that the cells are optimizing (regulating) the environmentally relevant features (\eg~the $\zcell$/$\ugcell$ yield).
That is, the culture observable state is an optimum, but only in the environmentally relevant ($\zcell,~\ugcell$) subspace (because the glucose-limited condition).
Once achieved such optimum, there are no further regulatory constraint affecting the degenerated dimensions.
An extra detail related with the experimental conditions, which supports such rationale, is that the studied cultures were run at small dilution rates ($D<0.5~\Dunits$).
This locates the cultures in a regime of slow growth rate (wild $\Ecoli$ can growth at $>2.2~\Dunits$~\cite{varmaStoichiometricFluxBalance1994}) and below the acetate switch~\cite{basanOverflowMetabolismEscherichia2015}.
This is relevant because the lower the growth rate, the less pressure is exerted on the cellular resources, and thus, internal regulations such as enzyme cost constraint (see Equation~\myref{eq:cost}) lose significance~\cite{begIntracellularCrowdingDefines2007, basanOverflowMetabolismEscherichia2015, vazquezMacromolecularCrowdingExplains2016}.
On the other hand, this also could be an explanation for the poor performance of $\FBA$ lineal formulations (see~\Figure{fig:corr_collage}).
$\FBA$ predicts the optimum at the ($\zcell$,~$\ugcell$) subspace, but there are no reason for the other dimensions to be also in an optimum.
This is the case, for example, of the known common ill-prediction of the flux through the glyoxylate cycle.
$FBA$ lineal formulations typically infer a zero flux (an optimum) where experiments report a non-zero value~\cite{demartinoIntroductionMaximumEntropy2018, rivas-astrozaMetabolicFluxConfiguration2020}.

\section{Conclusions}
	
To conclude, in this work, we exploit the Maximum Entropy Principle to provide a probabilistic description of the culture metabolism that can be used to infer the set of observable average fluxes, as well a description of the heterogeneity.
We introduce a new methodology to infer the metabolic properties of chemostat cultures, at steady state, under limiting nutrient conditions.
This inference problem was formulated and implemented for genome scale metabolic networks.
We showed that, at steady state and in limiting nutrient conditions, only two parameters are enough to capture all the relevant information contained on the data.
These parameters correspond to the two important constraints of the chemostat environment: one derived from the biomass mass balance, and the other from the limiting nutrient mass balance.
The technique was applied to a dynamical model of the chemostat, where the external conditions of the culture where linked with the internal cellular metabolism, and to experimental data from $\Ecoli$ cultures in a wide range of parameters.
Generally, our $\ME$ outperforms the inference obtained using different variants of $FBA$.

\section{Materials and methods}
\label{sec:Materials}

\subsection{E. coli continuous cultivation experimental data}
\label{sec:Ecoli_mat}

In order to test the predictive power of the different formulations, 
data of $\Ecoli$ glucose-limited continuous cultures was taken from literature.
Three different data sources where used: 
$\Kayser$~\cite{kayserMetabolicFluxAnalysis2005},
$\Nanchen$~\cite{nanchenNonlinearDependencyIntracellular2006} and
$\Folsom$~\cite{folsomPhysiologicalProteomicAnalysis2014}.

\subsection{E. coli metabolic network}

The metabolism of $\Ecoli$ was modeled using the metabolic network iJR904 (download link: \href{http://bigg.ucsd.edu/static/models/iJR904.mat}{bigg.ucsd.edu})~\cite{reedExpandedGenomescaleModel2003}.
The metabolic network was appropriately contextualized using the available experiment-specific data in the source publications.
If specific biomass composition data was available, the generic biomass equation in the metabolic network was also updated.
Additional enzymatic constraints was added according to~\cite{begIntracellularCrowdingDefines2007}.
For defining a bounded $\Vcell$ space, a few exchanges limits were added according to the largest values found at~\cite{varmaStoichiometricFluxBalance1994}.
Those bounds were not limiting in any of the studied experimental conditions.

\subsection{Implementation and software}

$\FBA$ was implemented using traditional linear programming and $\ME$ distributions were approximated using an adaptation of the $Expectation\ Propagation$ algorithm reported in~\cite{fernandez-de-cossio-diazMaximumEntropyPopulation2019}.
The implementation code can be found in GitHub at \href{https://github.com/josePereiro/Chemostat_EColi.jl}{\url{https://github.com/josePereiro/Chemostat_EColi.jl}}.
Follow the instructions for a complete reproduction of the results of this work.


\acknowledgments We are indebted with A. de Martino for useful discussions and with A. Muntoni for providing help with the implementation of the EP algorithm. The work was supported by the Horizon 2020 Marie Sk{\l}odowska-Curie Action-Rese
arch and Innovation Staff Exchange (MSCA-RISE) 2016 grant agreement 734439 (INFE
RNET: New algorithms for inference and optimization from large-scale biological 
data). It was also partially funded by the CITMA Project of the Republic of Cuba, PNCB-Statistical Mechanics of Metabolic Interactions-PN223LH010-015.

\bibliographystyle{unsrtnat}
\bibliography{MaxEnt_EColi.bib}

\clearpage

\section{Appendix}

\subsection{Chemostat dynamic simulation}
\label{sec:toy_dyn_app}

The toy network used on the dynamics comprehends the follow reactions: 
\begin{enumerate}
	\item \textit{glyc:} $(-1.0)Glc \rightarrow (2.0)Atp + (4.0)NADH + (1.0)AcCoa$
	\item \textit{ppp:} $(-1.0)Glc \rightarrow (2.0)NADH + (1.0)AcCoa$
	\item \textit{resp:} $(-2.0)NADH + (-1.0)Oxy \rightarrow (5.0)Atp$
	\item \textit{tac:} $(-1.0)AcCoa \rightarrow (1.0)Atp + (4.0)NADH$
	\item \textit{ferm:} $(-1.0)AcCoa \rightarrow (1.0)Atp + (1.0)Ac$
	\item \textit{ua:}   $\leftarrow (1.0)Ac$
	\item \textit{ug:}   $\rightarrow (1.0)Glc$
	\item \textit{uo:}   $\rightarrow (1.0)Oxy$
	\item \textit{atpm:} $(-8.4)Atp \rightarrow$
	\item \textit{z:}    $(-14.7)Glc + (-59.8)Atp \rightarrow$
\end{enumerate}

The biomass requirement was defined as $(Y_{X/Glc}) Glc~+~(GAM) Atp$ where $Y_{X/Glc} = 14.7 (\YXxunits)$ is the biomass/glucose yield~\cite{varmaStoichiometricFluxBalance1994} and $GAM = 59.8~(\YXxunits)$ is the growth associated $ATP$ maintenance demand~\cite{feistGenomescaleMetabolicReconstruction2007}.
The non-growth associated $ATP$ maintenance demand ($NGAM~= 8.4~(\YXxunits)$) \cite{feistGenomescaleMetabolicReconstruction2007} was modeled at the $atpm$ reaction.
All reaction are irreversible and open, except $atpm$ which both bounds were fixed to one.
The only limiting bound was at the glucose exchange ($\ugcell$) where $\ubg = 20~(\flxunits)$~\cite{varmaStoichiometricFluxBalance1994}, the rest was set to an arbitrary large number.
The model has two degree of freedom that we choose to be $\ugcell$ and $\zcell$.

We performed a dynamic simulation of the chemostat following equations~\myref{eq:cont_dyn_dX_dt} and \myref{eq:cont_dyn_dsg_dt}.
For computation, we discretize $\Vcell$ (and so $\Vmean$) using a quantum $\delta$ so:
\begin{gather}
	\ugcell \in \{ 0, 1\delta, 2\delta, ... \} \nonumber \\
	\zcell  \in \{ 0, 1\delta, 2\delta, ... \} \nonumber 
\end{gather}

Refactoring Eq.~\myref{eq:cont_dyn_dX_dt}, so we include flux and 
time discretization we have:
\begin{equation}
	\frac{\Delta X(\zcell, \ugcell)}{\Delta t} ~=~
	(1 - \epsilon) \zcell X(\zcell, \ugcell)
	+ \frac{\epsilon}{\Vcellvol_{\delta}} 
		\sum_{\zcellprime, \ugcellprime}^{\Vcell}  \zcellprime X(\zcellprime, \ugcellprime)
	- D X(\zcell, \ugcell)
	\label{eq:disc_dyn_dX_dt} 
\end{equation}
where $\Vcellvol_{\delta} \in \N$ is the total number of discrete 
regions contained at $\Vcell$.

Making a similar analysis, we can determine that the glucose concentration in the 
vessel evolves following:
\begin{equation}
	\frac{\Delta \sg}{\Delta t} = 
		- \sum_{\zcell, \ugcell}^{\Vcell} \ugcell X(\zcell, \ugcell) + (\cg - \sg)D
	\label{eq:disc_dyn_dsg_dt}
\end{equation}

Additionally, giving the values of $X(\zcell, \ugcell)$ we can compute a 
probability mass function:
\begin{equation}
	P(\zcell, \ugcell) = X(\zcell, \ugcell) / X
	\nonumber
\end{equation}
where $X = \sum_{\zcell, \ugcell}^{\Vcell} X(\zcell, \ugcell)$.

As stated on section \myref{sec:CH_dyn_sim}, equation \myref{eq:disc_dyn_dX_dt} and 
\myref{eq:disc_dyn_dsg_dt} are not sufficiently connected so the simulation
respects the implicit restriction of $\sg \ge 0$.
In order to achieve that we introduce a transformation over $P$ such:
\begin{equation}
	P^{\prime}(\zcell, \ugcell) = 
		\frac{P(\zcell, \ugcell) (\gamma - \ugcell/U_g)}
		{\sum_{\zcellprime, \ugcellprime}^{\Vcell} P(\zcellprime, \ugcellprime) (\gamma - \ugcellprime/U_g)}
\end{equation}
where $U_g$ is $\ug$ global maximum and $\gamma \in \R \wedge \gamma > 1$ is a parameter 
that ensures $\sum_{\zcell, \ugcell}^{\Vcell} \ugcell P^{\prime}(\zcell, \ugcell) \approx \cgDX$ at constant $X$.
Such transformation is applied over $P$ at every step of the simulation where $\sg \approx 0$
and $\sum_{\zcell, \ugcell}^{\Vcell} \ugcell P(\zcell, \ugcell) > \cgDX$ 
(when the moment inequality constraint is about to be broken).

\subsection{Unlimited culture dynamic}
\label{sec:unlimited_dyn_app}

\begin{figure}
	\centering
	\includegraphics[scale = 0.40]{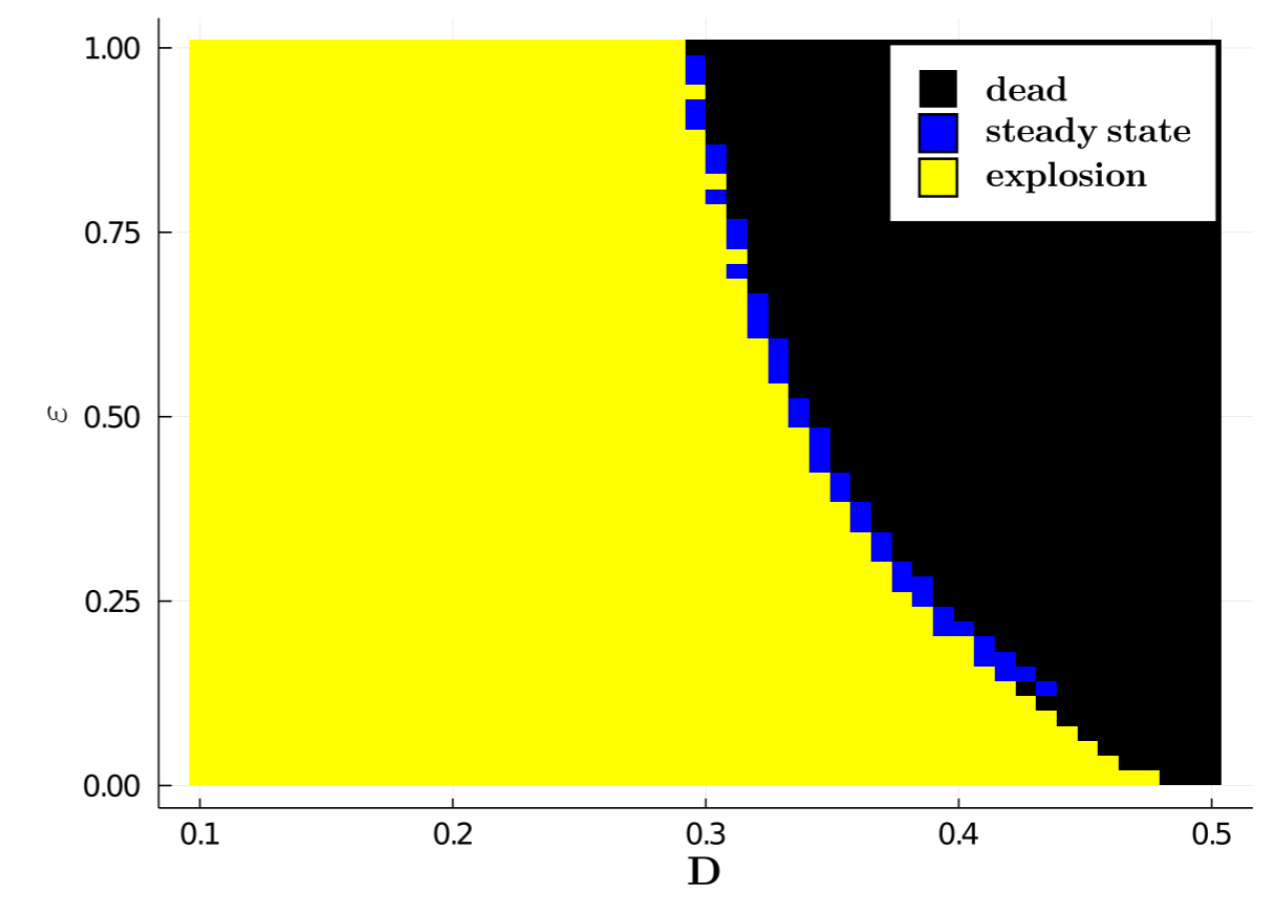}
	\caption{
		Heat map that represents the evolution of $X$ in a nutrient unlimited simulation ($\cg = +\infty$) as function of $D$ and $\epsilon$.
	}
	\label{fig:glc-unlim_study}
\end{figure}

In this section we study the dynamic of the chemostat when equations~\myref{eq:cont_dyn_dX_dt} and~\myref{eq:cont_dyn_dsg_dt} are decoupled.
This can be done by running simulations in a nutrient unlimited environment ($\cg = +\infty$).
In this case, no feedback is produced between the observables and the only significant environmental parameter will be the dilution rate.
The \Figure{fig:glc-unlim_study} shows a heat map with the $X$ value at the end of such simulations as a function of $D$ and $\epsilon$.
The simulations were stopped if either a non-trivial steady state condition was hit ($dX/dt \rightarrow 0$ and $X > 0$), the culture $X$ grows forever ($dX/dt > 0$ and $X > 10^6$) or the culture dies ($dX/dt < 0$ and $X < 10^{-6}$).
From the figure, the most relevant insight to be extracted is that the nutrient-unlimited dynamic typically lead to an unbounded growth or to a dead culture.
Non-trivial steady states are only possible at the interface between the two regions.
But, this interface represents an unstable regime. 
Small perturbations on either $D$ or $\epsilon$ will make such steady state unfeasible.
Is this tendency to increment $X$ which makes the culture to deplete all the nutrient once the limited condition is reestablished.
The culture nutrient uptake at steady state will be close to the input rate ($\ugmean \le \cgDX$).

\subsection{MaxEnt Algorithm}
\label{sec:ME_Alg_app}

The complete set of constraints defining $\Vcell$ (see equations \myref{eq:m_i_balance}, \myref{eq:v_j_bounds} and \myref{eq:cost}) can be expressed as:
\begin{gather}
	\Smat \vvec = \bvec \nonumber \\
	\lbvec \le \vvec \le \ubvec \nonumber
\end{gather}
where $\vvec \in \R^N$, $\lbvec \in \R^N$, $\ubvec \in \R^N$, $\Smat \in \R^{M \times N}$ and $\bvec \in \R^M$.

A uniform distribution mapped over $\Vcell$ can be written as~\cite{braunsteinAnalyticApproximationFeasible2017}:
\begin{equation}
	U(\vvec) \propto \delta(\Smat \vvec - \bvec)\prod_{n}^{N} \psi(v_n) \nonumber
\end{equation}
where $\delta(\Smat \vvec - \bvec)$ is a Dirac's delta with a non-zero value when $\vvec$ solves the linear system (encoding the exact constraints), and $\psi(v_n)$ is an indicator which equals one if $lb_n \le v_n \le ub_n$ and zero otherwise (encoding the relaxed constraints).

The extra constraints which define $\Vmean$ (see equations \myref{eq:Ch_ui_stst_ineq_constraints} and \myref{eq:Ch_mu_stst_constraints}) can be written as:
\begin{equation}
	\vmean \le \cvec \nonumber
\end{equation}
where $\vmean \in \Vmean$ and $\cvec \in \R^N$ is a constant.

Given that $\Vcell \subset \R^N$ is a convex space and the constraints over $\Vmean$ are linear, it can be proven that the distribution over $\Vcell$ which maximizes the entropy belong to the exponential family~\cite{jaynesInformationTheoryStatistical1957}.
Such exponential take the following form~\cite{fernandez-de-cossio-diazMaximumEntropyPopulation2019}:
\begin{equation}
	\PX(v~|~\betavec) \propto e^{(\betavec \vvec)} U(\vvec)
	\label{eq:Pexact}
\end{equation}
where the vector $\betavec \in \R^N$ contains the selection coefficients of each reaction flux in the network so $\vmean \le \cvec$ (where $\vmean = \int_{\Vcell} \vvec \PX(v~|~\betavec) d\vvec$) is meet, and the entropy is maximized.
A remark worth making is that the functional form of \myref{eq:Pexact} is generally intractable, so in this work, an approximated distribution obtained by $Expectation\ Propagation$ ($EP$) is used instead~\cite{braunsteinAnalyticApproximationFeasible2017}.
Such procedure is explained in details in the next section, and it is transparent for the current analysis.

In this work, we are trying to enforce two constraints over the mean values, and so, the model has two free parameters (two non-zero components in the $\betavec$ vector on \myref{eq:Pexact}).
One, $(\betaz)$, is used to restrict the average growth rate to equal the dilution rate $(\zmean = D)$, and the other, $(\betaug)$, is used restrict the average uptake of glucose in accordance with the glucose supply rate $(\ugmean \le c_g D / X)$.
Equation \myref{eq:Pexact} can be rewritten to make this more explicit:
\begin{equation}
	\PX(v~|~\betaz, \betaug) 
		\propto e^{(\betaz \zcell)} e^{(\betaug \ug)} U(\vvec)
		\nonumber
\end{equation}

Both moments $\zmean$ and $\ugmean$ depend on the selected values of $(\betaz, \betaug)$.
If the corresponding constraint is fulfilled the beta is called valid, $\betavz$ or $\betavug$ respectively.
Our goal is to find a pair of valid beta values, so the entropy is also maximal.
In order to do that we use the following algorithm:

\begin{algorithm}[H]
	\caption{ME Algorithm}
	\label{alg-me}
	\begin{algorithmic}[1]
		
		\Procedure{maxent2d}{}
			
			\State Init $\betaz$ and $\betaug$ at zero
			\State Compute $\zmean$ and $\ugmean$ using $\Pvbb{0}{0}$
			\If{Constraints $(\zeq)$ and $(\ugineq)$ are fulfilled}
				\State return $\betaz$ and $\betaug$ \label{mealg-ret1}
			\EndIf
			\\
			\State Update (grad. descend) $\betaz$ so constraint $(\zeq)$ is fulfilled
			\State Compute $\zmean$ and $\ugmean$ using $\Pvbb{\betaz}{0}$
			\If{Constraints $(\zeq)$ and $(\ugineq)$ are fulfilled}
				\State return $\betaz$ and $\betaug$ \label{mealg-ret2}
			\EndIf
			\\
			\While{Constraints $(\zeq)$ and $(\ugeq)$ are NOT fulfilled}
				\State Update (grad. descend) $\betaz$ so constraint $(\zeq)$ is fulfilled
				\State Update (grad. descend) $\betaug$ so constraint $(\ugeq)$ is fulfilled
				\State Compute $\zmean$ and $\ugmean$ using $\Pvbb{\betaz}{\betaug}$
			\EndWhile
			return $\betaz$ and $\betaug$ \label{mealg-ret3}
			
		\EndProcedure
		
	\end{algorithmic}
\end{algorithm}
where each beta update were performed using a simple gradient descent till the given target was approximated.

As can be noticed, the entropy is not explicitly maximized in any of the gradient descents.
But, the algorithm ensures that each returned pair $(\betavz, \betavug)$ do specify the distribution with the maximal entropy from all valid ones.
Indeed, the above algorithm is nothing but standard maximization of entropy (following~\cite{fernandez-de-cossio-diazCellPopulationHeterogeneity2019}), with the only peculiarity that we must also deal with inequality constraints on average values of the distribution, such as $\vmean_i \le C$ for some flux $i$.
The above algorithm is based on the idea that if this constraint is not satisfied automatically when one solves the MaxEnt problem without including it, then the optimal solution (when considering also this constraint), will satisfy instead the equality constraint $\vmean_i = C$.
The proof states as follows:

\begin{proof}
	Let $\PX(v)$ be a distribution over fluxes $v \in \Vcell$.
	The entropy:
	\begin{equation}
		S[\PX] = - \int_{\Vcell} \ln(\PX(\vvec)) \PX(\vvec) d\vvec \nonumber
	\end{equation}
	is a concave functional of $\PX$.
	Let $\Pspace$ be any convex space of probability distributions.
	For instance, $\Pspace$ can be the space of probability distributions with support $\Vcell$.
	We are interested in finding the solution of a $\ME$ problem, of the form:
	\begin{gather}
		\text{maximize arg}_{\PX \in \Pspace} ~ S[\PX] \nonumber \\
		\text{subject to:} \nonumber \\
		\vmean_i \le a_i \nonumber
	\end{gather}

	We denote by $\PX^{c}$ and $\vmean^{c}$ ($c$ stand for constrained) the resulting distribution and its average vector.
	Additionally, we define $\PX^{g}$ and $\vmean^{g}$ to be the solution of the problem if we ignore the inequality constraint.
	Clearly $S[\PX^{g}] \ge S[\PX^{c}]$ because $S[\PX^{c}]$ has the additional inequality constraint ($g$~stand for global maximum).
	If $\vmean^{g}_i \le a_i$, both problems have the same solution, that is, $\PX^{c} = \PX^{g}$ and $\vmean^{c} = \vmean^{g}$ (which is the case on lines \myref{mealg-ret1} and \myref{mealg-ret2} on the algorithm \myref{alg-me}).

	If $\vmean^{g}_i > a_i$, the two solutions necessarily differ.
	We show that in this case $\vmean^{c}_i = a_i$ necessarily.
	Suppose, to the contrary, that $\vmean^{c}_i < a_i$.
	This means that $\PX^{c}$ is a local optimum of the entropy within $\Pspace$.
	However, since the entropy is concave and $\Pspace$ is a convex space, then $\PX^{c}$ must also be a global optimum, that is, $\PX^{c} = \PX^{g}$.
	But then we have a contradiction, $a_i < \vmean^{g}_i = \vmean^{c}_i < a_i$.
	Therefore, $\vmean^{c}_i < a_i$ is impossible, and we must have $\vmean^{c}_i = a_i$, as stated (which is the case for the line \myref{mealg-ret3} on the algorithm \myref{alg-me}).

\end{proof}

\subsection{Expectation Propagation} 
\label{sec:EP_app}

As stated in the last section, a $\ME$ distribution directly derived from the definition of $\Vcell$ and $\Vmean$ has the form:
\begin{equation}
	\Pexact(\vvec) 
		\propto e^{(\betavec \vvec)} \delta(\Smat \vvec - \bvec)\prod_{n}^{N} \psi(v_n)
		\nonumber
\end{equation}

Through Gaussian elimination, we can transform the matrix $\Smat$ to a row echelon form:
\begin{equation}
	\Smat \equiv \left[ \identmat \vert \Gmat \right] \nonumber
\end{equation}
where $\identmat \in \R^{M \times M}$ is an identity matrix and $\Gmat \in \R^{M \times (N-M)}$.

The structure of the linear constraint induced by the row echelon representation suggests splitting the $\vvec$ variable vector into two sets of variables: the first M variables (dependent) and a second set of N-M variables (independent).
To do so, we define:
\begin{equation}
	\vvec \equiv (\vdepvec, \vindvec) \nonumber
\end{equation}
where, as we said, $\vdepvec \in \R^{M}$ and $\vindvec \in \R^{N-M}$ and
\begin{equation}
	\vdepvec = \bprimevec - \Gmat\vindvec \nonumber
\end{equation}
where $\bprimevec \in \R^{M}$ is the transformed (after Gaussian elimination) version of $\bvec$.

We rewrite the probability density function in terms of the new variable definitions:
\begin{equation}
	\Pexact(\vdepvec, \vindvec) \propto 
		e^{(\betaindvec \vindvec)}
		e^{(\betadepvec \vdepvec)}
		\delta(\identmat \vdepvec + \Gmat \vindvec - \bprimevec)
		\prod_{m}^{M} \psi(\vdepvec_m)
		\prod_{n}^{N - M} \psi(\vindvec_n)
		\nonumber
\end{equation}

We now can compute the $\vindvec$ marginal as:
\begin{equation}
	\Pexact(\vindvec) \propto  
		\int \big\lbrack
				e^{(\betaindvec \vindvec)}
				e^{(\betadepvec \vdepvec)}
				\delta(\identmat \vdepvec + \Gmat \vindvec - \bprimevec)
				\prod_{m}^{M} \psi(\vdepvec_m)
				\prod_{n}^{N - M} \psi(\vindvec_n)
			\big\rbrack d\vdepvec
	\nonumber
\end{equation}

Note that the delta makes this integral to have a single non-zero contribution at $\vdepvec = \bprimevec - \Gmat\vindvec$, so it solves to:
\begin{equation}
	\Pexact(\vindvec) \propto 
		e^{(\betaindvec \vindvec)}
		e^{\betadepvec (\bprimevec - \Gmat \vindvec)}
		\prod_{m}^{M} \psi(\bprimevec_m - [\Gmat\vindvec]_m)
		\prod_{n}^{N - M} \psi(\vindvec_n)
	\label{eq:Pexact-ind}
\end{equation}

The indicators priors $\psi$ makes the marginals of this distribution hard to compute, so we instead use the approximate multivariate Gaussian $\phi(\vvec; \avec, \dvec)$ with mean vector $\avec \equiv (\adepvec, \aindvec)$ and variance vector $\dvec~\equiv~(\ddepvec, \dindvec)$ to formulate an approximated join distribution:
\begin{equation}
	\Papprox(\vindvec) \propto 
		e^{(\betaindvec \vindvec)}
		e^{\betadepvec (\bprimevec - \Gmat \vindvec)}
		\phi(\bprimevec - \Gmat \vindvec; \adepvec, \ddepvec)
		\phi(\vindvec; \aindvec, \dindvec)
	\label{eq:Papprox-ind}
\end{equation}
which is a multivariate Gaussian distribution that can be expressed in standard form as:
\begin{gather}
	\Papprox(\vindvec) \propto 
		\exp \biggl[ 
				(\vindvec - \vmeanindvec)^T {\Sigmaindmat}^{-1} (\vindvec - \vmeanindvec)
			\biggl]
	\nonumber \\
	\Sigmaindmat = (\Dindmat + \Gmat^T \Ddepmat \Gmat)^{-1}
	\nonumber \\
	\vmeanindvec = \Sigmaindmat(
		\Gmat^T \Ddepmat (\bprimevec -  \adepvec) +
		\Dindmat \aindvec -
		\Gmat^T \betadepvec + \betaindvec
		\nonumber
	)
\end{gather}
where $\Ddepmat \in\R^{M \times M}$ and $\Dindmat \in\R^{(N - M) \times (N - M)}$ are the priors' covariance matrices where all covariances are zero and the diagonals equals $1/\ddepvec$ and $1/\dindvec$ receptively.

The parameters of the dependent variables are easily derived from the independents as:
\begin{gather}
	{\Sigmadepmat} = \Gmat {\Sigmaindmat} \Gmat^T \nonumber \\
	\vmeandepvec = \bprimevec - \Gmat \vmeanindvec \nonumber
\end{gather}

Now, we are in conditions to apply $Expectation\ Propagation$ as describe in \cite{braunsteinAnalyticApproximationFeasible2017} to find the parameters $\avec$ and $\dvec$ of the Gaussian priors that better approximate \ref{eq:Papprox-ind} to \ref{eq:Pexact-ind}.

\subsection{Study of additional biases}
\label{sec:Bias_app}

As discussed in the \Section{sec:MaxEnt}, an advantage of $\ME$ over $\FBA$ is that it uses more effectively all the information contained in the constraints, which allows inferring properties of the culture metabolic state other than the observable flux configuration.
But, it also makes $\ME$ more sensible to the introduction of unnoticed biases.
To gain a deeper insight, we replicate the same analysis over the $\Ecoli$ experimental data using the $\ME$ formulation described at~\cite{fernandez-de-cossio-diazMaximumEntropyPopulation2019}.
The new model (called in this section $\ME^1$) uses a single beta parameter (the super index account for the number of non-zero $\beta$ parameters).
The only difference with our model (called $\ME^2$ in this section) is that the nutrient limiting constraint is simplified from $\ugmean \le \cgDX$ to $\ugcell \le \cgDX$.

\begin{figure}
	\centering
	\includegraphics[scale = 0.1]{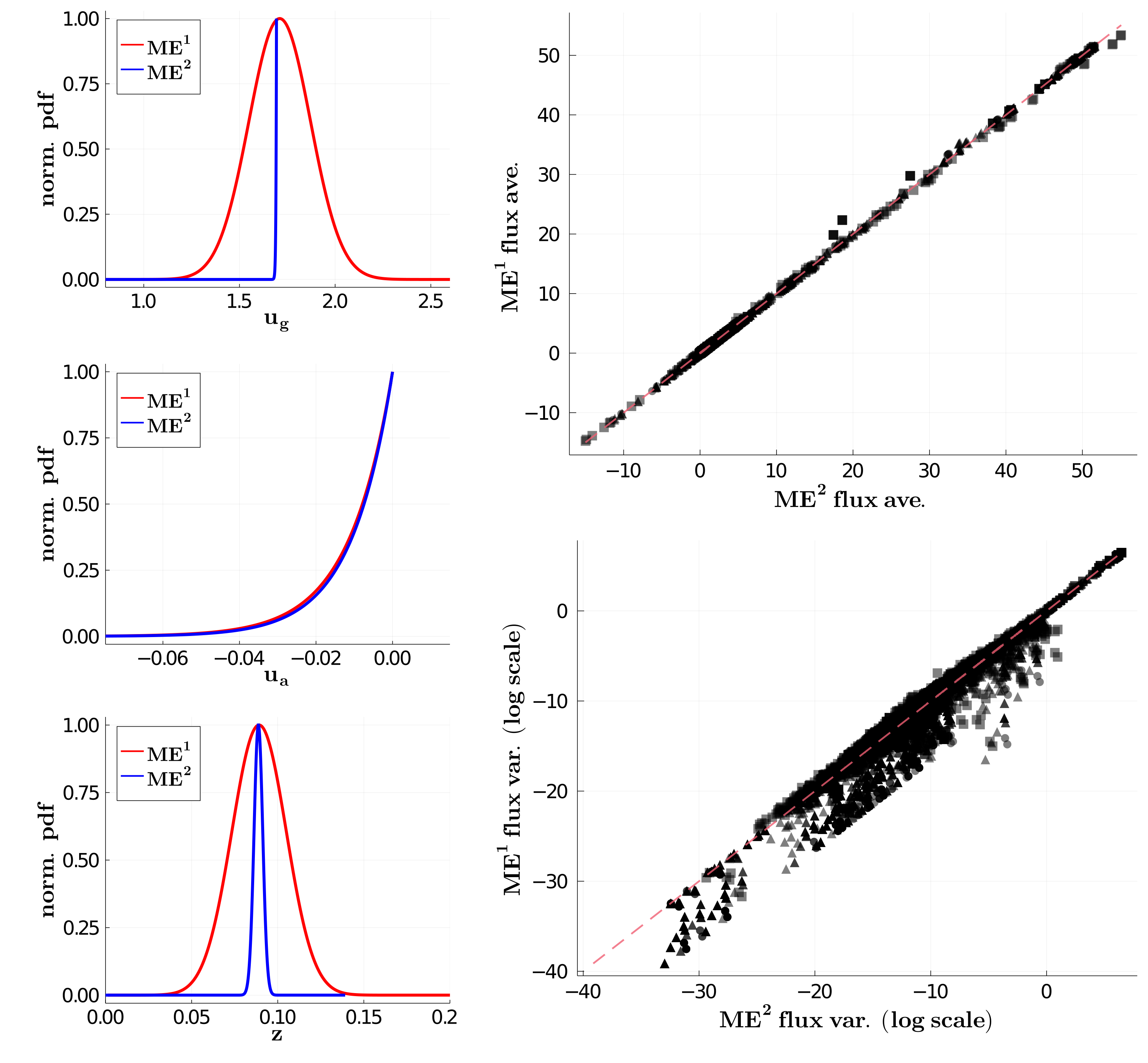}
	\caption{
		Study of the effect of the different $\ME$ formulations on the inferred flux distributions.
		The left column shows selected marginals of the glucose uptake flux, $\ug$, the acetate production, $u_a$, and the biomass production rate, $\zcell$, for both $\ME$ formulations in an experiment (rep. 4) from $\Nanchen$~\cite{nanchenNonlinearDependencyIntracellular2006}.
		In the right column, it is shown the correlations of all flux averages (top) and all flux variances between both formulation for all data sources.
	}
	\label{fig:ME_unbiased_vs_biased}
\end{figure}

\Figure{fig:ME_unbiased_vs_biased} presents the comparison between both formulations.
In the left column of the figure we show the marginal distributions (for one experimental condition) for three selected fluxes.
The first marginal (top-left) is the one corresponding to the uptake of glucose.
As mentioned before, the codification of the chemostat constraint of this flux
is the only difference between the two formulations.
As can be seen, both marginals differ substantially.
This subtle difference, to consider that the knowledge of an observable restricts $\Vcell$, is sufficient to produce a major difference in the solution of $\ME$ ($\eg$ the heterogeneity of the culture).
Additionally, because the network imposes a structural constraint that is
reflected in a correlation between the fluxes, this discrepancy is propagated
to others.
This can be noticed in the marginal of the biomass reaction (bottom-left).
In both cases the reduction of $\Vcell$ in $\ME^1$ formulation resulted in 
distributions with smaller degeneracy.
A large-scale study of such an effect is shown in the right column of the same figure.
There, we show a comparison between both formulations averages (top-right) and variances (bottom-right) for all the fluxes in all experimental conditions.
As can be seen, the averages are not particularly affected, but the variances
(which are shown in a log scale) are consistently smaller for the $\ME^1$ formulation.

Although $\ME^2$ is more rigorous in this sense, this formulation might be not totally free from biases associated with the exchanges.
For instance, equation~\myref{eq:Ch_ui_stst_ineq_constraints} shows that a metabolite not present in the feed medium should have a negative or zero average exchange rate, which means that the culture can only potentially produce it, not consume it.
Even though this is an observable constraint, we made the assumption that ($c_i = 0 \implies u_i \le 0$).
As mentioned before, the correct formal methodology for encoding such observable constraints is by moving its corresponding components in the $\betavec$ vector so that the selected $\PME$ does complaint with the restrictions and~$\Vcell$ stays properly unaffected.
Because this needs to be done for all the metabolites that the network might produce, it would increase the number of free non-zero $\betavec$ components that needs to be tuned for inferring the $\PME$ distribution, which would make its computation more challenging.

An example of such phenomena can be appreciated for the acetate exchange rate, whose marginal is shown in the left column of the figure (middle panel).
The acetate exchange marginal is abruptly cut at zero.
This provokes that, in case of any degeneration, its average gets a value greater than zero.
This depends on the assumption that cells can not consume acetate, which is not directly derived from any constraint imposed by the chemostat.
As discussed before $\ME$ predicts a wrong non-zero acetate production rate (see \Figure{fig:corr_collage} top-right panel).
This might be another possible cause of discrepancy with the experiments.

\end{document}